\documentclass[twocolumn]{aastex62}

\usepackage{newtxtext,newtxmath}
\usepackage[T1]{fontenc}
\usepackage{ae,aecompl}
\usepackage{graphicx}	
\usepackage{amsmath}	
\usepackage{amssymb}	

\defcitealias{martin_2013}{M13}
\defcitealias{ibata_2014b}{I14}

\hypersetup{
  colorlinks=true,
  urlcolor=blue,
  citecolor= blue,
  linkcolor = blue
}

\graphicspath{{./}{figures/}}

\shorttitle{\textsc{Auriga2PAndAS}}
\shortauthors{Thomas et al.}

\begin{document}

\title{Observing the stellar halo of Andromeda in cosmological simulations: the \textsc{Auriga2PAndAS} pipeline}

\correspondingauthor{Guillaume F. Thomas}
\email{guillaume.thomas@nrc-cnrc.gc.ca}

\author[0000-0002-2468-5521]{Guillaume F. Thomas}
\affiliation{NRC Herzberg Astronomy and Astrophysics, 5071 West Saanich Road, Victoria, BC, V9E 2E7, Canada}
\affiliation{Instituto de Astrof\'isica de Canarias, E-38205 La Laguna, Tenerife, Spain}
\affiliation{Universidad de La Laguna, Dpto. Astrof\'isica, E-38206 La Laguna, Tenerife, Spain}

\author[0000-0002-1349-202X]{Nicolas F. Martin}
\affiliation{Universit\'e de Strasbourg, CNRS, Observatoire astronomique de Strasbourg, UMR 7550, F-67000 Strasbourg, France}
\affiliation{Max-Planck-Institut f\"ur Astronomie, K\"onigstuhl 17, D-69117, Heidelberg, Germany}

\author{Azadeh Fattahi}
\affiliation{Department of Physics, Institute for Computational Cosmology, Durham University, South Road, Durham, DH1 3LE, UK}

\author[0000-0002-3292-9709]{Rodrigo A. Ibata}
\affiliation{Universit\'e de Strasbourg, CNRS, Observatoire astronomique de Strasbourg, UMR 7550, F-67000 Strasbourg, France}

\author{John Helly}
\affiliation{Department of Physics, Institute for Computational Cosmology, Durham University, South Road, Durham, DH1 3LE, UK}
\affiliation{Carnegie Observatories, 813 Santa Barbara Street, Pasadena, CA 9110}

\author{Alan W. McConnachie}
\affiliation{NRC Herzberg Astronomy and Astrophysics, 5071 West Saanich Road, Victoria, BC, V9E 2E7, Canada}

\author{Carlos Frenk}
\affiliation{Department of Physics, Institute for Computational Cosmology, Durham University, South Road, Durham, DH1 3LE, UK}

\author{Facundo A. G\'omez}
\affiliation{Instituto de Investigaci\'on Multidisciplinar en Ciencia y Tecnolog\'ia, Universidad de La Serena, Ra\'ul Bitr\'an 1305, La Serena, Chile}
\affiliation{Departamento de Astronom\'ia, Universidad de La Serena, Av. Juan Cisternas 1200 Norte, La Serena, Chile}

\author[0000-0001-9667-1340]{Robert J.J. Grand}
\affiliation{Max-Planck-Institut f\"ur Astrophysik, Karl-Schwarzschild-Str. 1, D-85748 Garching, Germany}

\author{Stephen Gwyn}
\affiliation{NRC Herzberg Astronomy and Astrophysics, 5071 West Saanich Road, Victoria, BC, V9E 2E7, Canada}

\author{Dougal Mackey}
\affiliation{Research School of Astronomy and Astrophysics, Australian National University, Canberra, ACT 2611, Australia}

\author[0000-0003-3816-7028]{Federico Marinacci}
\affiliation{Department of Physics \& Astronomy, University of Bologna, via Gobetti 93/2, I-40129 Bologna, Italy}

\author[0000-0003-3308-2420]{R\"udiger Pakmor}
\affiliation{Max-Planck-Institut f\"ur Astrophysik, Karl-Schwarzschild-Str. 1, 85748 Garching, Germany}

\begin{abstract}
We present a direct comparison of the Pan-Andromeda Archaeological Survey (PAndAS) observations of the stellar halo of M31 with the stellar halos of 6 galaxies from the Auriga simulations. We process the simulated halos through the {\sc Auriga2PAndAS} pipeline and create PAndAS-like mocks that fold in all observational limitations of the survey data (foreground contamination from the Milky Way stars, incompleteness of the stellar catalogues, photometric uncertainties, etc). This allows us to study the survey data and the mocks in the same way and generate directly comparable density maps and radial density profiles. We show that the simulations are overall compatible with the observations. Nevertheless, some systematic differences exist, such as a preponderance for metal-rich stars in the mocks. While these differences could suggest that M31 had a different accretion history or has a different mass compared to the simulated systems, it is more likely a consequence of an under-quenching of the star formation history of galaxies, related to the resolution of the {\sc Auriga} simulations. The direct comparison enabled by our approach offers avenues to improve our understanding of galaxy formation as they can help pinpoint the observable differences between observations and simulations. Ideally, this approach will be further developed through an application to other stellar halo simulations. To facilitate this step, we release the pipeline to generate the mocks, along with the six mocks presented and used in this contribution. 
\end{abstract}


\section{Introduction}

How were L$\star$-galaxies like the Milky Way or Andromeda formed? Behind this relatively simple question hides a very complex picture, starting from cosmological scales and trickling down to sub-parsec scales. In the standard $\Lambda$CDM cosmological paradigm, the formation of structures happens hierarchically, with dark matter halos growing though a succession of mergers, their inhabitant galaxies absorbing the smaller galaxies located at the center of these accreted halos \citep{searle_1978,white_1978,moore_1999}. Due to dynamical friction, a large fraction of the accreted stars, coming mostly from major mergers, sinks rapidly toward the Galactic disk \citep[e.g.][]{boylan-kolchin_2008,pillepich_2015,gomez_2017}, where dynamical timescales are of order a few hundred millions years, removing the traces of these accretion events. Fortunately, the imprints of the merger history are conserved over a very long period of time \citep[many Gyr;][]{bullock_2005,johnston_2008} in the most distant, and faintest (reaching surface brightnesses of $\mu_V \sim 35$ mag.arcsec$^{-2}$), component of a galaxy, its stellar halo. Therefore, this component is key to unveiling the details of the formation history of L$\star$-galaxies, such as the number of past accretion events, the mass of the accreted galaxies, and the epoch of these events. The stellar halos of Milky Way-like galaxies are very diffuse and extended, reaching out to their virial radius \citep[e.g.][]{helmi_1999,cooper_2010,gomez_2013}. They are often complex structures, very inhomogeneous and clumpy \citep[e.g.][]{bell_2008,martinez-delgado_2010,duc_2015,merritt_2016,mcconnachie_2018}, with the presence of numerous satellite galaxies, globular clusters, and their tidal debris (streams, shells, plumes, etc.) visible for several Gyrs \citep{johnston_2008}. 

Because stellar halos are very extended and only account for a few percent of the light emitted from a galaxy, it is very challenging to observe them. Despite earlier studies that detected faint components around nearby galaxies and diskovered several stellar streams \citep[e.g.][]{malin_1997,mihos_2005,martinez-delgado_2010,martinez-delgado_2013}, it is only recently that integrated light photometry can efficiently be used to measure the surface brightness profiles of galactic halos \citep{duc_2015,merritt_2016,merritt_2020,trujillo_2016,dsouza_2018,wang_2019}. A tremendous amount of work has been invested to better characterize the point spread function (PSF) of photometric observations and to improve the optics of the instruments, allowing for a better separation of the scattered light from galaxies and contaminating scattered light produced by galactic cirrus, stars, and other compact objects \citep{slater_2009}. These relatively inexpensive observations\footnote{In comparison to observations that aim to resolve individual stars.} allow for the observation of a great number of galaxies with a great diversity of halos profiles stemming from the stochasticity of the galaxy formation process. Yet, with this technique, it is very challenging to study the detailed properties of individual halos, such as their radial age and metallicity distributions or, even harder, to obtain their dynamical properties.

On the other hand, resolved stellar photometry is more informative, and allows to measure the physical properties of individual halos, at the cost of very expensive observations. In particular, the Milky Way halo has been extensively studied due to its relatively close distance that allows main sequence stars to be observed out to a few dozen kpc \citep[e.g.][]{carollo_2007,carollo_2010,ivezic_2008,juric_2008,bell_2008,bell_2010,watkins_2009,sesar_2011,deason_2012,gomez_2012,deason_2013,xue_2011,xue_2015,ibata_2017b,fukushima_2018,fukushima_2019,hernitschek_2018,thomas_2018a}. However, it is extremely costly to study the halo of the Milky Way, since a large area of the sky has to be observed and, ideally, in a broad range of photometric bands to facilitate the separation of different stellar populations. Furthermore, to study the profile of the halo of our Galaxy, it is necessary to use three-dimensional distances due to our central position, but only a few percent of the halo stars have distances known with good enough accuracy (\citealt{ibata_2017b}, but see \citealt{thomas_2019a}). Thus, in some respects, the halos of nearby external galaxies are easier to study than the halo of our Galaxy. They cover a smaller area on the sky and, with some assumptions, it is possible to use the projected distances to infer their profile with a better accuracy and at larger distances than for the Milky Way. 

In this endeavor, the Hubble Space Telescope (HST) has been a powerful tool to study the halo of nearby galaxies, in particular thanks to the ACS Nearby Galaxy Survey Treasury \citep[ANGST]{dalcanton_2009} and to the Galaxy Halos, Outer
disks, Substructure, Thick disks, and Star clusters \citep[GHOSTS][]{radburn-smith_2011} programs. These surveys have revealed a great diversity of masses, of metallicity distributions, and of stellar populations in the stellar halos of L$\star$-galaxies that otherwise have similar disk morphologies, masses and luminosities \citep[e.g.][]{ibata_2009,monachesi_2016,harmsen_2017}. However, these surveys are often pencil-beam and do not offer a complete view of the halo of the target galaxies. 

Therefore, the halo of the Andromeda galaxy (M31) is an important and unique object due to our ability to probe the stellar populations of an L$\star$-galaxy at very faint stellar magnitudes, as done with HST \citep[e.g.][]{brown_2006,brown_2007,richardson_2009} and, in parallel, over a very large areas, as done by the Pan-Andromeda Archaeological Survey \citep[PAndAS,][]{mcconnachie_2009,mcconnachie_2018}. This latter survey, covering $>400$ square degrees around M31, has been extremely important to study the morphology and the chemistry of the stellar halo of a galaxy similar to the Milky Way \citep[i.e.][]{ibata_2014b}, but also to quantify its level of substructures \citep[stellar streams, dwarf galaxies, or globular clusters; i.e.][]{mackey_2010,collins_2013,lewis_2013,martin_2013,huxor_2014,mcconnachie_2018}. This survey has shown, for instance, that the Andromeda galaxy has a stellar halo that is about 15 times more massive than that of the Milky Way \citep[$\sim 10^{10}$ M$_\odot$ for M31 and $4-8 \times 10^{8}$ M$_\odot$ for the MW;][]{bell_2008,deason_2011,deason_2019,ibata_2014b}, with a shallower profile slope, and displays a metallicity gradient that drops from $\langle[$Fe/H$]\rangle=-0.7$ at 30 kpc to $\langle[$Fe/H$]\rangle=-1.5$ at 150 kpc \citep[][hereafter \citetalias{ibata_2014b}]{ibata_2014b}, contrary to what is observed in the Milky Way \citep[e.g.][]{ivezic_2008,juric_2008,sesar_2011,xue_2015,ibata_2017b}. These differences between the two galaxies tend to indicate that they have been subjected to two very different formation history \citep[e.g.][]{deason_2013,gilbert_2014,harmsen_2017,dsouza_2018}.

From a simulation point of view, analytical models and dark-matter only simulations (with different prescriptions to include baryons) that take into account only the accreted component of a halo showed very early that the broad diversity of halos observed in nearby galaxies is a natural consequence of the stochasticity of the merger/accretion history for each galaxy \citep[e.g.][]{bullock_2005,renda_2005,font_2006,purcell_2007,delucia_2008,cooper_2010}. Moreover, these simulations also demonstrated that the majority of the halo material of a given galaxy has been contributed by the few most massive satellite galaxies accreted through its history \citep[e.g.][]{robertson_2005,cooper_2010,deason_2016,amorisco_2017}. However, as shown by \citet{bailin_2014}, the simplifying assumptions made by dark-matter only simulations tend to a systematic underestimate of the halo concentration and an incorrect quantification of the level of substructure of the halo. More recent cosmological hydrodynamical simulations such as Eris \citep{pillepich_2015}, APOSTLE \citep{sawala_2016,oman_2017}, Auriga \citep{grand_2017}, Illustris TNG \citep{pillepich_2018a}, Latte/FIRE \citep{wetzel_2016,hopkins_2018}, or ARTEMIS \citep{font_2020} can model more representative stellar halos by self-consistently including the baryonic (stellar) distribution. With these simulations, it has been possible to confirm that a correlation exists between the number of significant progenitors, the metallicity and the mass of a halo. For instance, they show that halos made by a few significant progenitors tend to be more massive, more concentrated, and with a significant negative metallicity gradient \citep{deason_2016,dsouza_2018,pillepich_2018,monachesi_2019}. In addition, these simulations include an {\it in-situ} stellar component of the stellar halo, composed of stars born in the galactic disk and that have been ejected by the interactions with sub-haloes or molecular clouds. They also include stars formed in streams of gas stripped from infalling satellites and dominate the inner halo of L$\star$ galaxies \citep[e.g.][]{purcell_2010,font_2011,pillepich_2015,cooper_2015}.

Therefore, important information on the formation history of galaxies like ours can be gained by comparing the simulations to the observations. The observations are useful to constrain the assumptions and limitations of the simulations, and the simulations can provide useful physical context to explain the formation of the galaxies. However, only a rigorous apples-to-apples comparison between the simulations and the observations, using the same tools, the same methods, with consistent biases and limitations, and with the same assumptions, are meaningful to improve the underlying physical models used by the simulations, especially concerning the baryonic physics.

The aim of this paper is to pursue a direct comparison between simulations and observations of Andromeda's stellar halo, by transforming 6 galaxies from the {\sc Auriga} simulations into PAndAS-like mocks, as presented in Section \ref{sec_method}. The mocks are compared to the observations using the same tools in Section \ref{sec_results}, including a discussion on the importance of the {\it in-situ} population in the simulations in Section \ref{sec_insitu}. The implications of the results are analysed and discussed in Section \ref{sec_analyse}, and we conclude in Section \ref{conclusion}.

\section{Method} \label{sec_method}

In this section, we describe how we transform the simulated stellar halos of 6 of the 30 Andromeda-like galaxies from the Auriga suite of simulations \citep{grand_2017} to ``realistic'' stellar halo mocks as if they were observed by the Pan-Andromeda Archaeological Survey \citep[PAndAS,][]{mcconnachie_2009,mcconnachie_2018}. 

The PAndAS survey was a Large Program of the Canada-France-Hawaii Telescope (CFHT) that observed the surrounding of the Andromeda (M31) and Triangulum (M33) galaxies. This program comprises 406 fields of $1 \degr \times 1\degr$ obtained in the $g$ and $i$ bands with the MegaPrime/MegaCam camera between 2008 and 2011 and also includes fields from a pilot survey between 2003 and 2008 with the same instrumental set up. A detailed description of the PAndAS data, their acquisition, reduction, and the resulting catalogues of resolved stars can be found in \citet{ibata_2014b} and \citet{mcconnachie_2018}. In particular, the location of the survey fields can be seen in Figure 1 of \citetalias{ibata_2014b}.

The {\sc Auriga} simulations are a suite of thirty cosmological magneto-hydrodynamical zoom-in simulations of MW-like galaxies, made with the moving mesh magnetohydrodynamics code {\sc Arepo} \citep{springel_2010,pakmor_2016}. These galaxies were selected from the parent dark matter only cosmological simulation EAGLE \citep{theeagleteam_2017} to have a similar halo mass than the MW and to satisfy being isolated at $z = 0$. We refer the reader to \citet{grand_2017} for a detailed description of these simulations. 

\subsection{Generation of the initial mock stellar catalogues} \label{method}

The mock stellar catalogues used as inputs of this pipeline are computed from the Auriga simulations \citep{grand_2017} in a very similar manner to what was previously presented by \citet{grand_2018a} to produce the {\sc Aurigaia} mock stellar catalogues. To generate these stellar mocks, two methods were presented, the \textsc{HITS-mocks} and the \textsc{ICC-mocks}. These two methods differ in how the ``stars,'' generated from the stellar particles of the simulations are distributed in phase space, as well as by the choice of stellar evolution models. For the rest of this paper, we use exclusively the \textsc{ICC-mocks}, based on \citet{lowing_2015}. The reason is that this method produces smoother distributions of ``stars'' and avoids discrete clumps at the coordinates of the parent stellar particles, while still preserving the phase-space distribution. Therefore, with this method, the presence of artificial features, such as fake clumps, in the distribution of ``stars''\footnote{In the rest of the paper, the mock stars generated from a simulation's star particles will be referred as ``stars'' to differentiate them from the real stars of the PAndAS observations.} from the stellar halo in the Auriga simulations is minimized. With this method, each parent stellar particle is split into $N$ ``stars,'' assuming a Chabrier IMF, and drawing from a model stellar population of a given age and metallicity according to the \textsc{parsec} isochrones \citep{bressan_2012}, which take into account the evolution of the massive end of the IMF for stellar particles with old ages. The stellar parameters and the absolute magnitudes in the CFHT $g$ and $i$ bands\footnote{Here, we use the pre-2014 CFHT/MegaCam photometric system, before the current set of filters was built. It is worth noting that the fields obtained before 2007 have been observed with a slightly different $i$-band filter \citep[see Figure 3 of][]{mcconnachie_2018} but this change is subtle enough that it should not affect our results. } are computed for each mock ``star'' from the \textsc{parsec} isochrones \citep{bressan_2012}, using the age and metallicity of the parent stellar particle.

The foreground MW extinction was added at a letter step, as will be described later in Section \ref{sec_A2P}. Indeed, our study is focused on the region of M31 observed by the PAndAS survey, i.e. the stellar halo, where the extinction caused by the interstellar medium (ISM) of M31 is negligible. The large majority of the dimming of stars observed in the stellar halo of M31 is due to the foreground extinction caused by the ISM of the MW, and is included later in the pipeline. 

\begin{figure*}
\centering
  \includegraphics[angle=0, viewport= 0 0 1025 695, clip,width=17.0cm]{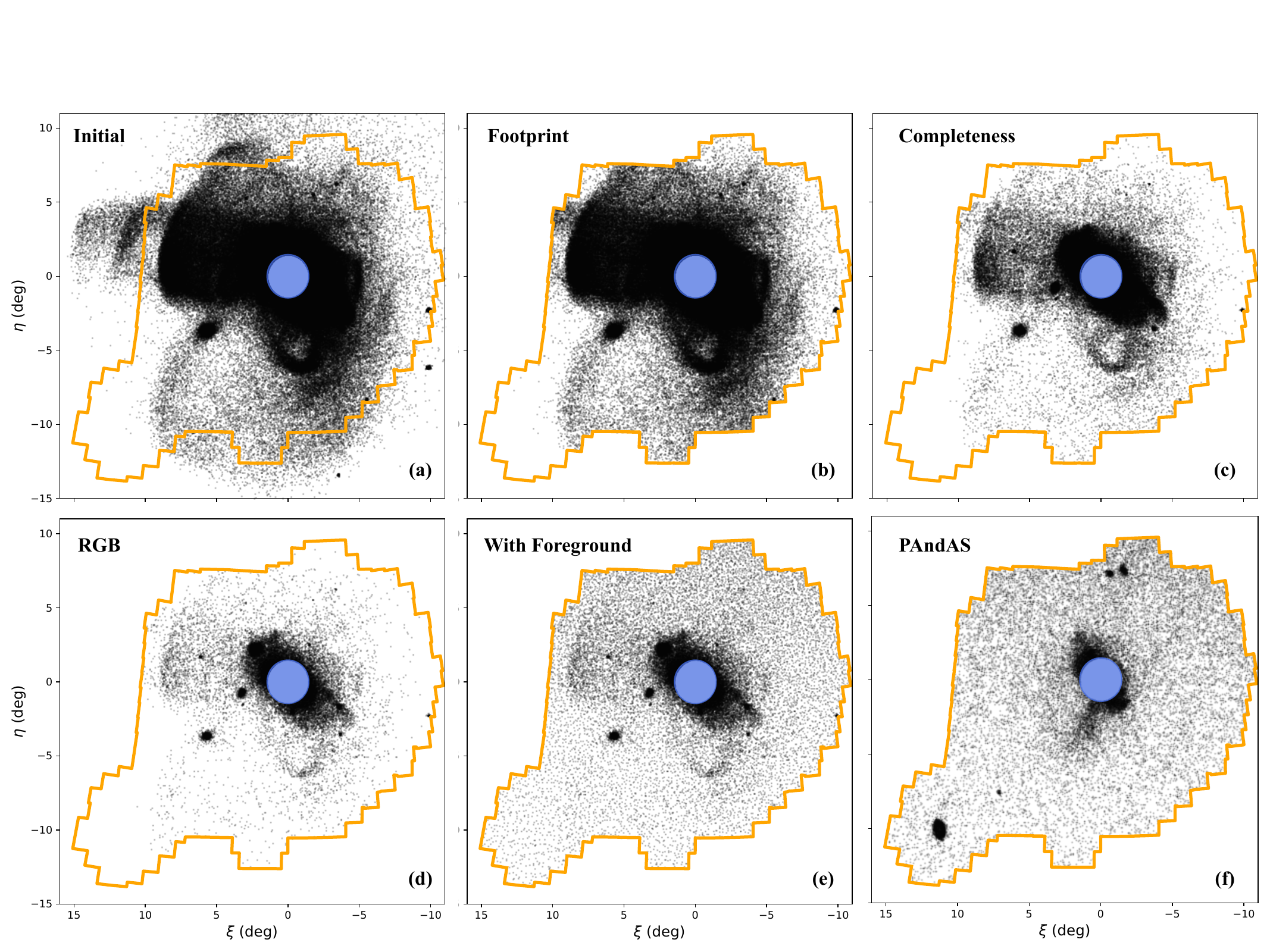}
   \caption{Example of the projection of $1,000,000$ initial ``stars'' from halo H23 into the observational space. In all panels, we show the plane of the sky tangent to the celestial sphere at the location of M31. Panel (a) presents the initial distribution of mock ``stars''. Panel (b) shows the same stars after application of the PAndAS footprint. In panel (c), we further apply the completeness criteria, while panel (d) shows the selection of ``stars'' within the Red Giant Branch box of \citet{martin_2013}. Finally, panel (e) displays the final state of the mock, after adding ``stars'' from the Milky Way foreground contamination model. In all panels, the orange polygon traces the external border of the PAndAS footprint and the blue circle highlights the inner 20 kpc that have been initially removed from the mock to mask the majority of the disk stars in the simulations since they are  irrelevant to this study. Panel (f) show an a sub-selection in PAndAS that have a similar density stars than in panel (e) outside $6\degr$ from the center of M31 and 2$\degr$ outside M33. The grey scale is the same in all panels.} 
\label{steps}
\end{figure*}

It is important to note here that for the analysis of the simulations, detailed in Section \ref{sec_results}, the central projected 20 kpc (or $1.47\degr$ at the chosen distance of M31) are not taken into account. This region is severely incomplete in the PAndAS survey and a study of this region would require dedicated work \citep[e.g. HST PHAT survey][]{dalcanton_2012}. Therefore, from this stage and for each mock, we decided to completely avoid ``stars'' within the a central sphere of 20 kpc radius. This action considerably reduces the size of the mocks and the computation time to apply the \textsc{Auriga2PAndAS} pipeline since the large majority of the simulation ``stars'' reside in this region. By applying this cut to the three-dimensional distances instead of the projected distances at this step, we keep the possibility to project the galaxy using a random point of view around the galaxy (see below).

\subsection{The \textsc{Auriga2PAndAS} pipeline} \label{sec_A2P}

From these raw mocks that include all ``stars'' down to an absolute magnitude of $M_i=2$ ($\sim 6$ magnitude fainter than the tip of the RGB at the distance of M31), we degrade the data so as to be as close as possible to the observed PAndAS data. The different steps we perform are:
\begin{enumerate}
    \item placing the simulation at the distance of M31, orienting them similarly to M31 and projecting it on the sky;
    \item masking out ``stars'' that are not in a PAndAS field or behind saturated foreground stars;
    \item computing the apparent magnitude of the ``stars'' and their associated uncertainties;
    \item making the data incomplete following the observed, field-specific completeness functions;
    \item selecting the RGB ``stars'' with a color-magnitude cut;
    \item adding the contamination from foreground MW stars and background unresolved galaxies.
\end{enumerate}

The effects of these steps on the distribution of ``stars'' for one of the halos is visible in Figure \ref{steps}. We now discuss each of these stages in turn.

\subsubsection{Projection}
Once the initial mock stellar catalogues are built, we project them on the sky, placing the center of the simulated galaxy at the position of M31 (R.A., Dec)=($00^h42^m44.330^s$, $+41\degr 16'07.50"$) \citep{skrutskie_2006} and at a heliocentric distance of 778 kpc \citep{conn_2011,conn_2012}. The simulations are oriented such that their Galactic disks have a similar orientation to the disk of the Andromeda galaxy, with an inclination $i = 77.5\degr$ and a position angle $\theta = 37.7\degr$, following \citet{metz_2007}. Even though throughout this paper we only show and analyze a single realization of each simulated galaxy, the {\sc Auriga2Pandas} pipeline allows the user to rotate the disk around the vertical galactic axis with a random angle for each realization, so as to allow for future work to study statistically the properties of a given ``observed'' simulation with different points of view. The coordinates of each ``star'' are computed in the plane tangential to the celestial sphere at the location of M31, $(\xi, \eta)$. As per convention, $\xi$ increasing toward the west and $\eta$ toward the north. Throughout the rest of the paper, only one realization for each Auriga halo is considered. A statistical analysis of the simulations, incorporating different points of view for each halo will be performed in the future.

The apparent velocities of the ``stars'' are also computed from their physical velocity, to which we add the global motion of M31, ($V_{los}$, $\mu_\alpha^*$, $\mu_\delta$)=(-300 km.s$^{-1}$, 65 $\mu$as.yr$^{-1}$, -57$\mu$as.yr$^{-1}$) \citep{mcconnachie_2012,vandermarel_2019}, assuming a Solar radius of 8.1 kpc \citep{gravitycollaboration_2018a}, a circular velocity at the Solar radius of 229 km.s$^{-1}$ \citep{eilers_2019} and a Solar peculiar motion of (U$_\odot$, V$_\odot$, W$_\odot$) = (11.1, 12.24, and 7.25) km.s$^{-1}$ in local standard of rest coordinates \citep{schonrich_2010}.

In panel (a) of Figure \ref{steps}, we show the impact of this step to mock H23. Here, we only show a random sub-sample of $1,000,000$ ``stars'' in the figure for clarity.

\begin{figure}
\centering
  \includegraphics[angle=0, clip,width=7.5cm]{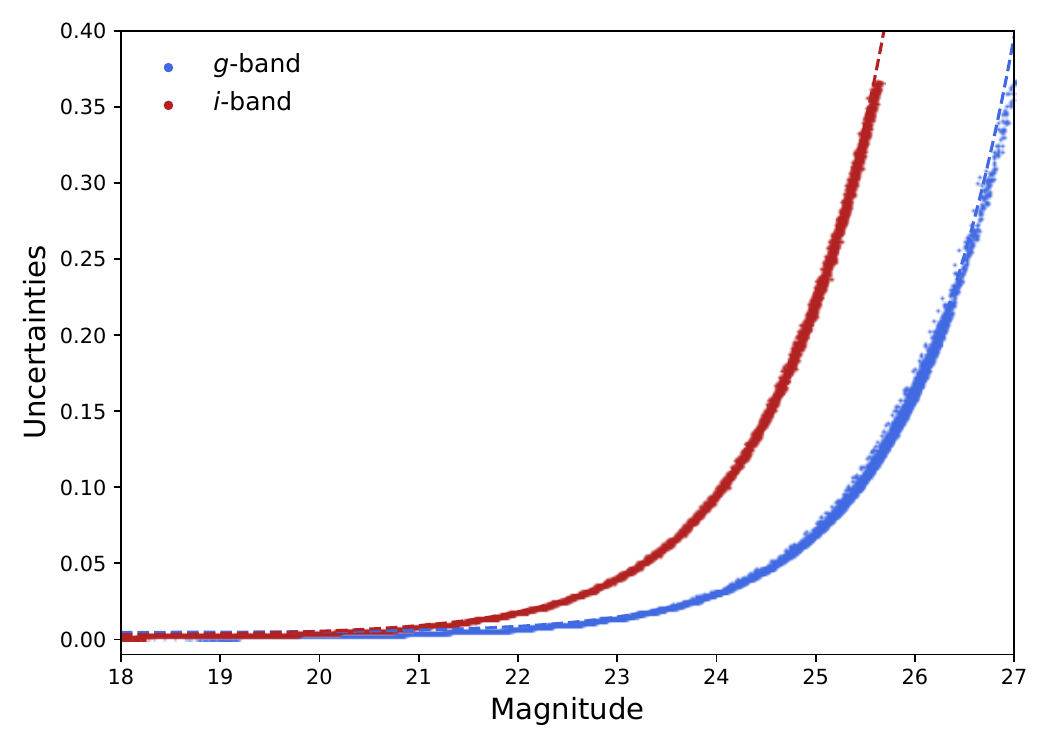}
   \caption{Photometric uncertainties as a function of the observed magnitude in the $g$ (blue) and $i$ bands (red) for the point sources observed in reference field 10. The fitted relations of equation (2) are represented by the two dashed lines.}
\label{uncertainties}
\end{figure}

\begin{figure}
\centering
  \includegraphics[angle=0, clip, viewport= 145 0 575 720,width=8.5cm]{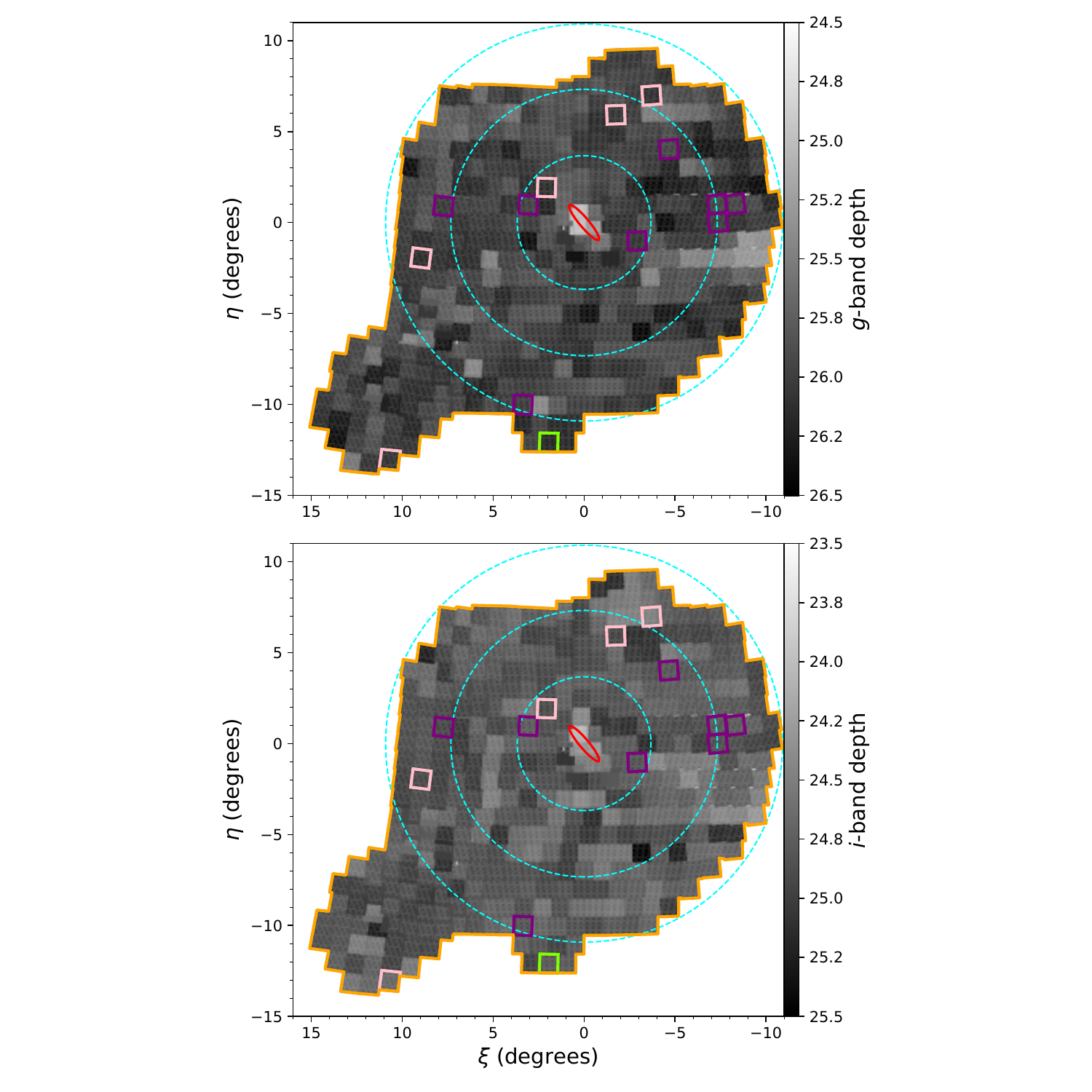}
   \caption{Map of the photometric depths of all PAndAS fields for point sources in the $g$ (top) and $i$ bands (bottom). The cyan circles correspond to projected distance of 50, 100, and 150 kpc and the red ellipse shows the approximate edge of the main stellar disk of M31. The reference field used to calibrate the photometric uncertainties is represented by the green square. The fields around the globular clusters used to calibrate the completeness function are highlighted by the purple squares, while the pink squares show the fields used to validate this calibration.}
\label{depth}
\end{figure}

\subsubsection{PAndAS footprint mask}
We apply the mask of the PAndAS coverage to the mocks to remove ``stars'' outside the footprint. This includes the ``stars'' that are outside the external border of the survey, but also ``stars'' that fall in the small number of holes between some of the observed fields or in the gaps between the lines of CCDs in the MegaPrime/MegaCam camera \citep[see Figure 2 of][]{mcconnachie_2018}. ``Stars'' at locations in the ($\xi$, $\eta$) plane that correspond to bright foreground stars in the PAndAS data are also removed from the mock catalogue,similar to how we treat the observed data (see section 2 of \citealt{ibata_2014b}). Panel (b) of Figure \ref{steps} shows the impact of this step of the procedure on the 1,000,000 stars of mock H23 shown in panel (a) of the same figure.

\subsubsection{Apparent magnitude}
For all unmasked ``stars'', their apparent magnitudes in the $g$ and $i$ bands are determined from the absolute magnitudes, given in the initial mock stellar catalogue, and from their individual heliocentric distances, computed in the previous step. To account for the foreground extinction produced by the ISM of the MW, the absolute magnitudes are reddened using the $E(B-V)$ extinction map of \citet{schlegel_1998}. We further assume a 10\% uncertainty on these values to mimic our likely imperfect extinction correction for the PAndAS data and to avoid reddening the data by the exact same amount we will later de-redden them by when studying the mocks like we study the PAndAS data. With the knowledge of $E(B-V)$ at a given location in the survey, we redden the data, using the coefficients from \citet{martin_2013} (hereafter referred as \citetalias{martin_2013}), such that
\begin{equation}
\begin{array}{c}
g = g_0 +3.793\ E(B-V)\\
i = i_0 + 2.086\ E(B-V).
\end{array}
\end{equation}
\noindent Here, $g_0$ and $i_0$ refer to the perfect apparent magnitude of a ``star'' contained in the mock catalogue, while $g$ and $i$ are their reddened equivalent, comparable to the observed and calibrated magnitudes in the PAndAS catalogue.

\begin{figure}
\centering
  \includegraphics[angle=0, clip,width=8.5cm]{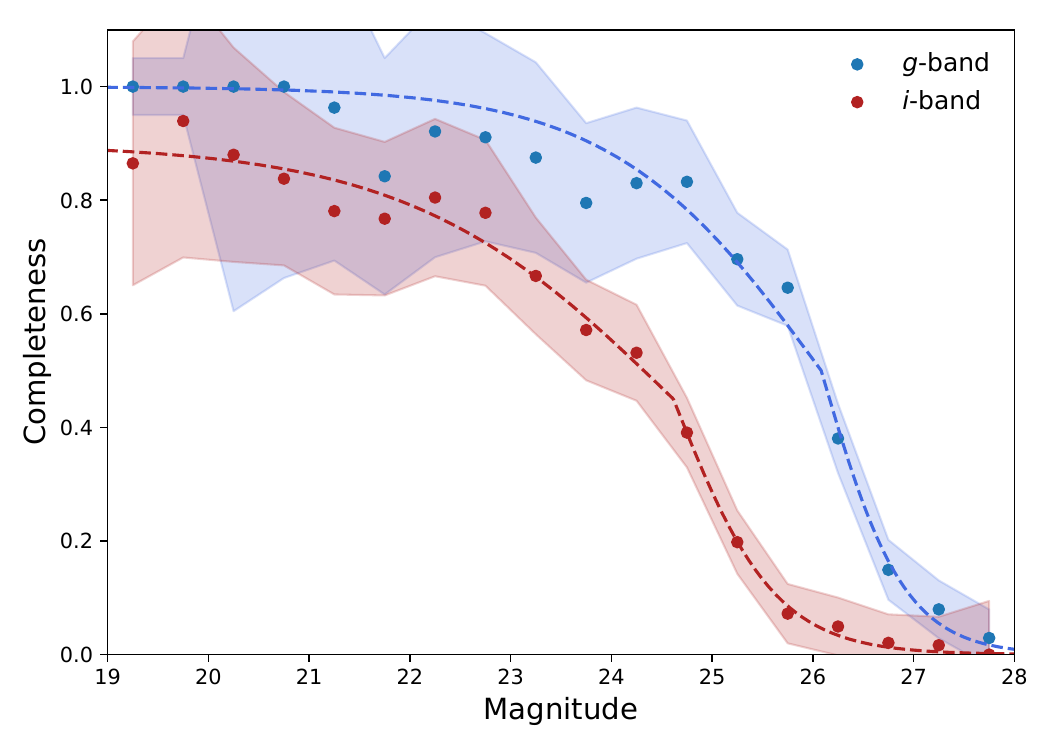}
   \caption{Completeness in the $g$ (blue) and $i$ bands (red) of the PAndAS survey, built from the data in reference fields 35, 229, 243, 261, 263, 267, 274, and 335. The shaded area represents the uncertainties on the completeness values, including a constant value of 0.05 added to the Poissonian uncertainties. The dashed lines show the best fit relations detailed in Eq. \ref{eq_comp}. }
\label{completeness}
\end{figure}

The photometric uncertainties that we assign to each mock ``star'' are then computed from their apparent magnitude. In order to build a model for the uncertainties as a function of magnitude, we first chose an observed reference field (field 10), for which we isolate all point sources\footnote{As in \citetalias{ibata_2014b}, we define as point sources objects that have a classification flag from the Cambridge Astronomical Survey Unit (CASU) pipeline \citep{irwin_2001} of either -1 or -2  in both the $g$ and $i$ bands.}. We use those to build a model of the photometric uncertainties as a function of magnitude (Figure~\ref{uncertainties}) that we model with the following functions:

\begin{equation}
\begin{array}{c}
\displaystyle
\delta g =0.032 \ \exp\left(\frac{g-24.25}{1.10}\right)+0.004\\
 \\
\displaystyle
\delta i =0.112 \ \exp\left(\frac{i-24.25}{1.13}\right)+0.003.
\end{array}
\label{eq_uncert}
\end{equation}

\noindent However, it is important to note that the depth, and so the related photometric uncertainties, is different for each field, due to changes in the observing conditions of any PAndAS field. These depths, $lim_{g,j}$ and $lim_{i,j}$, as defined by the magnitude for which the observed photometric uncertainties reach 0.2 mag, are determined for all PAndAS fields and shown in Figure~\ref{depth} for the two observed bands. Substituting $g-(lim_{g,j}-lim_{g,10})$ for $g$ (respectively $i-(lim_{i,j}-lim_{i,10})$ for $i$) in equation~\ref{eq_uncert} allows us to shift the uncertainty models from reference field 10 to any field $j$. Only moderate inhomogeneities are apparent over the survey footprint, and only the two fields covering the M31 disc are noticeable outliers, with a shallower depth in both the $g$ and $i$ bands.

For every ``star'' in the mock, we determine the PAndAS field it falls in given its $(\xi,\eta)$ location and we randomly draw an uncertainty in $g$ and in $i$ based on the models described above. We then update the apparent magnitude of this ``star'' by adding random Gaussian deviates based on these modeled $\delta g$ and $\delta i$. The ``noisy'' measurements are those stored in the mock catalogues.

Finally, the $g$ and $i$ magnitudes are corrected from the foreground extinction using the exact $E(B-V)$ from \citet{schlegel_1998}.

\subsubsection{Completeness}
With the next step, we aim to take the photometric completeness of the survey into account. To estimate the completeness of PAndAS in each band, we compare the number of PAndAS point source objects as a function of magnitude with the number of point sources observed in the deeper Hubble Space Telescope (HST) fields obtained by \citet[]{mackey_2007} and \citet[]{mackey_2013} around globular clusters of M31. These deep HST fields have been observed in the $F606W$ and $F814W$-bands\footnote{The Hubble $F606W$ and $F814W$ filters correspond roughly to $V$ and $I$ in the Johnson system, so we will abusively refer to them as such in the rest of the paper.} by the Advanced Camera for Surveys (ACS), and are $\sim$2.5--3 magnitudes deeper than the PAndAS survey. The data reduction of these fields and the star/galaxy separation will be described in a future contribution (Mackey et al. in prep.). 

Over the 46 fields observed by HST, we selected the 14 fields located around globular clusters B517, H1, PA~02,  PA~03,  PA~06, PA~11, PA~18, PA~43, PA~44, PA~45, PA~46, PA~47, PA~49, and PA~56. These HST fields have been selected because they are located in PAndAS fields, for which the depth is similar in the $g$ and $i$ bands, namely in fields 35, 229, 243, 261, 263, 267, 274, and 335, with a mean depth in these field of $26.03\pm{0.03}$ in $g$ and $24.85\pm{0.04}$ in $i$, close to the mean depth of the overall PAndAS survey. The transformation from the HST photometric system to the MegaPrime/MegaCam system are performed using objects detected as point sources in the $V$ , $I$, $g$ and $i$ bands in these 14 fields. It is worth noting here that the objects located in the inner 25 arcsec of each clusters are not taken into account in the rest of the analysis as these regions can suffer from crowding. By cross-matching the HST fields to the PAndAS ones, we find that the relations between these two photometric systems for point sources are the following:

\begin{equation}
\begin{array}{l}
g = \begin{cases} -0.18+0.98\ (V-I) + V, & \mbox{if } (V-I)<1.30 \\ 0.92+0.13\, (V-I) + V, & \mbox{if } (V-I)\geq 1.30 \end{cases}\\
\\
i = 0.33+ 0.18\ (V-I) +I.
\end{array}
\end{equation}

Using these color transformations, the completeness of PAndAS in those fields is determined independently in $g$ and $i$, by comparing the number of objects identified in PAndAS as point sources, in this specific band, per bin of $0.5$ mag, to the number of point sources objects identified in $V$ and $I$ in the HST fields. This method was preferred over an artificial star-test experiment directly into the PAndAS images because this latter method is very computationally expensive and would not have improved the quality of our work or of the pipeline, especially regarding other steps where the assumptions made have stronger impacts on the final representation of the simulations (e.g. the split of stellar particles into "stars"). The assumption made by the method chosen here is that all stars down to the PAndAS depth are present in the HST observations, which seems reasonable since the HST fields are $\sim 2.5-3$ magnitudes deeper than the PAndAS observations. The completeness of the $g$ and $i$ bands determined in this way are shown in Figure \ref{completeness}. We find that the completeness functions, $C_g(g)$ and $C_i(i)$ for the $g$ and $i$ bands, respectively, can be fit reasonably with the following functions:

\begin{equation}
\begin{array}{l}
\displaystyle
C_g(g)= \begin{cases}
\displaystyle
\left[ 1+ \exp\left( \frac{g-26.08}{1.04}\right)\right]^{-1}, & \mbox{if } (g<26.08) \\
\displaystyle
 \left[ 1+ \exp\left( \frac{g-26.08}{0.41}\right)\right]^{-1}, & \mbox{if } (g \geq 26.08) \end{cases}\\
\\
\displaystyle
C_i(i)= \begin{cases}
\displaystyle
0.9\ \left[ 1+ \exp\left( \frac{i-24.62}{1.31}\right)\right]^{-1}, & \mbox{if } (i<24.62) \\
\displaystyle
0.9\ \left[ 1+ \exp\left( \frac{i-24.62}{0.50}\right)\right]^{-1}, & \mbox{if } (i \geq 24.62). \end{cases}
\end{array}
\label{eq_comp}
\end{equation}

\begin{figure}
\centering
    \includegraphics[angle=0, clip,width=8.5cm]{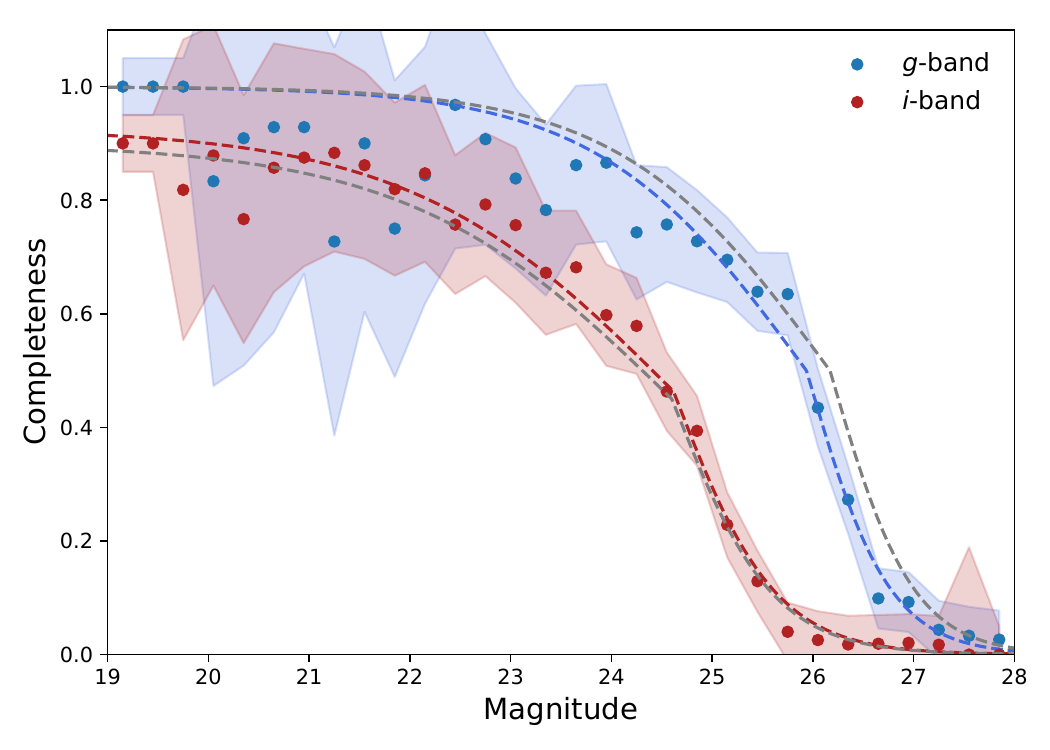}
   \caption{Completeness of the PAndAS fields around dwarf galaxies And X, And XVII, And XXI, And XXIV, And XXV, and And XXVI. The grey lines show the fits of the models described in Eq.(\ref{eq_comp}), shifted to the depth of the fields while the red and blue lines show the best fit of the completeness for these fields. The shaded areas represent the 1-$\sigma$ of the completeness.}
\label{completeness_dsphs}
\end{figure}

As for the photometric uncertainties, the completeness of a large survey like PAndAS that has been observed over many years varies from field to field, reflecting the specific observational conditions of each field. To account for this spatial variation, it is possible to replace $g$ and $i$ in Eq.(\ref{eq_comp}) by $g'=g-(lim_{g,j}-26.03)$ and $i'=i-(lim_{i,j}-24.85)$, respectively, where $lim_{g,j}$ and $lim_{i,j}$ are the depth of field $j$ in the $g$ and $i$ bands found previously, and $26.03$ and $24.85$ are the mean depths of the reference fields for the completeness determination. We validated this method with additional HST fields located around the dwarfs galaxies And X, And XVII, And XXI, And XXIV, And XXV and And XXVI \citep{martin_2017}. These HST fields are located in fields 3, 194, 296, 376, and 390, for which the mean depth is $26.12\pm 0.15$ in $g$ and $24.83\pm{0.14}$ in $i$. The results for these fields are presented on Figure \ref{completeness_dsphs}, with the grey lines showing the completeness determined by our method, and the blue/red dashed lines representing the best fit of the completeness in this field in $g$ and $i$. Our method gives similar results to the best fit, with only differences of a few percent, especially for $i<23.5$ and for $g<25.5$, where RGB stars at the distance of Andromeda are present \citepalias{martin_2013}.

Once we have a completeness model, we apply this model to the mocks by using an acceptance-rejection method such that ``stars'' are removed from the catalogue if $Rand(0,1)>C_g(g,lim_{g,j})$ and  $Rand(0,1)>C_i(i,lim_{i,j})$. Here, $Rand(0,1)$ is a number drawn from a uniform distribution between 0 and 1. The result of this operation yields the ``star'' distribution shown in panel (c) of Figure \ref{steps} for halo H23.

\begin{figure*}
\centering
  \includegraphics[angle=0, clip, viewport= 0 0 960 770 ,width=14.5cm]{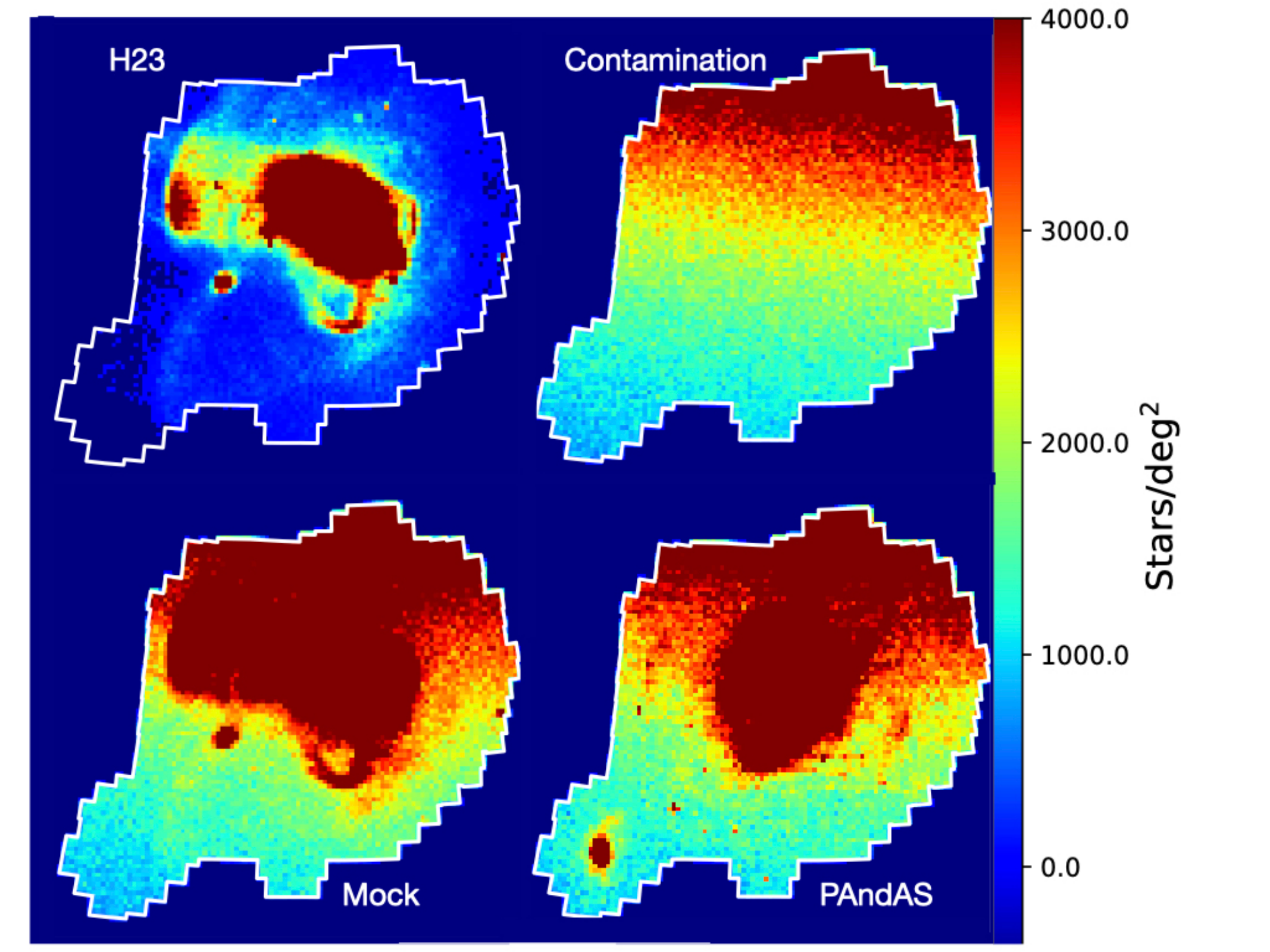}
   \caption{Upper left: density of remaining particles from the H23 stellar mock in the M31 RGB box at the end of the pipeline. Upper right: density of contaminant particles, integrated over the M31 RGB box. Lower left: total density of particles of the final "observed" mock over the M31 RGB box. Lower right: density of actual observed objects in the M31 RGB box over the PAndAS  footprint.}
\label{density_RGB}
\end{figure*}

\subsection{Selection of the RGB stars and inclusion of the foreground contamination} \label{sel_rgb}
Following \citetalias{martin_2013}, we keep stars that have a color and magnitude compatible with those of M31 RGB stars and remove ``stars'' outside of the selection box whose $(i_0, (g-i)_0)$ vertices are $(21, 0.7), (21, 2.3), (23.5, 0.4), (23.5, 1.6)$. For halo H23, this step yields the distribution shown in panel (d) of Figure \ref{steps}.

\begin{figure*}
\centering
  \includegraphics[angle=0, clip ,width=17.5cm]{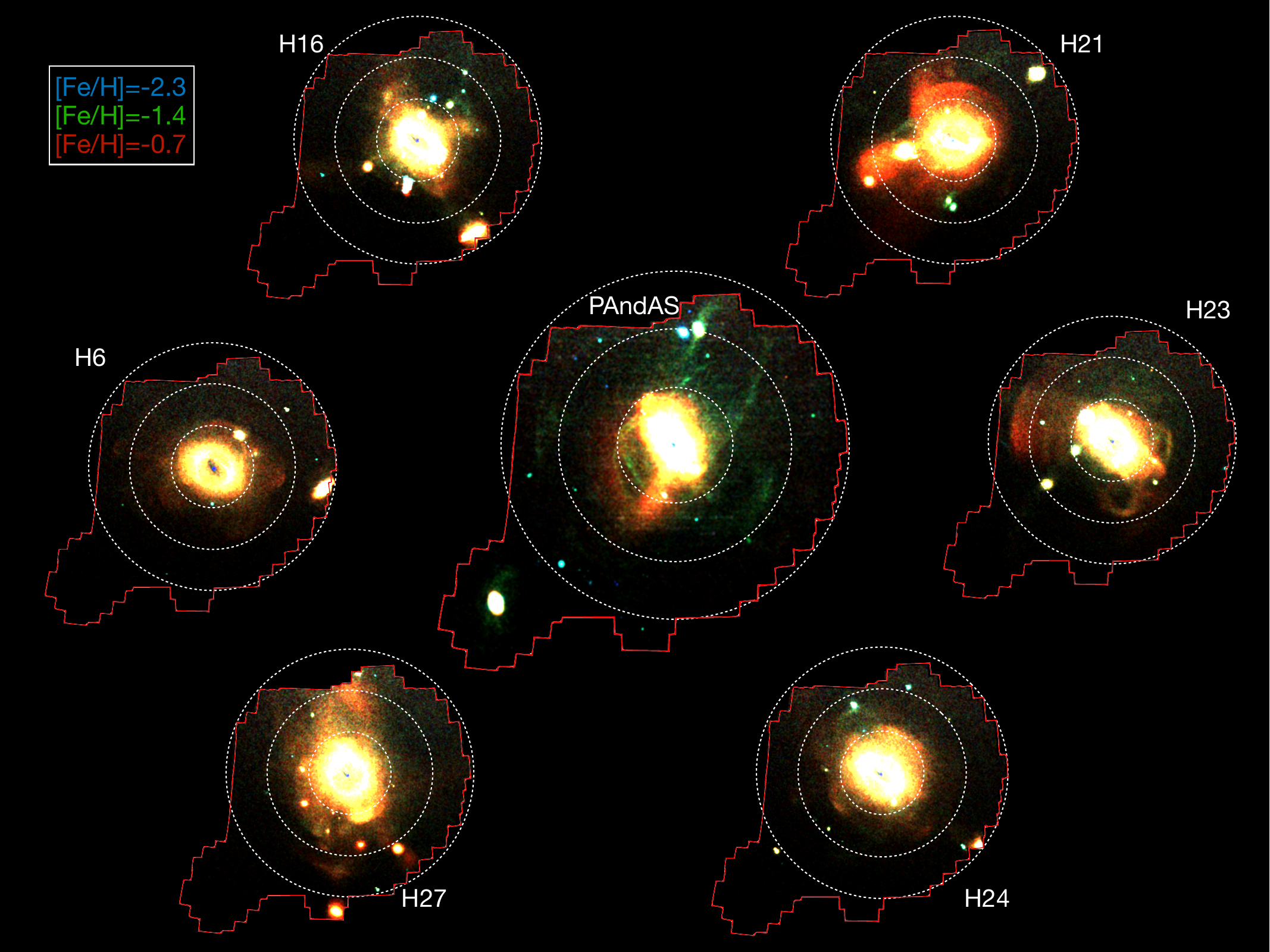}
   \caption{Red-green-blue maps of the observed PAndAS survey (at the center) and of the 6 \textsc{Auriga} mock simulations. Each color channel is a matched-filter map using, as the signal, the CMD of an old RGB population at the distance of M31 with metallicities of $[\mbox{Fe/H}]=-2.3$ (blue),  $[\mbox{Fe/H}]=-1.4$ (green), and $[\mbox{Fe/H}]=-0.7$ (red), and with a local background CMD that follows the contamination model of \citetalias{martin_2013}. The white circles correspond to projected distances of 50, 100, and 150 kpc.}
\label{MF_all}
\end{figure*}

\begin{figure*}
\centering
  \includegraphics[angle=0, clip, viewport= 0 15 1025 630 ,width=17.5cm]{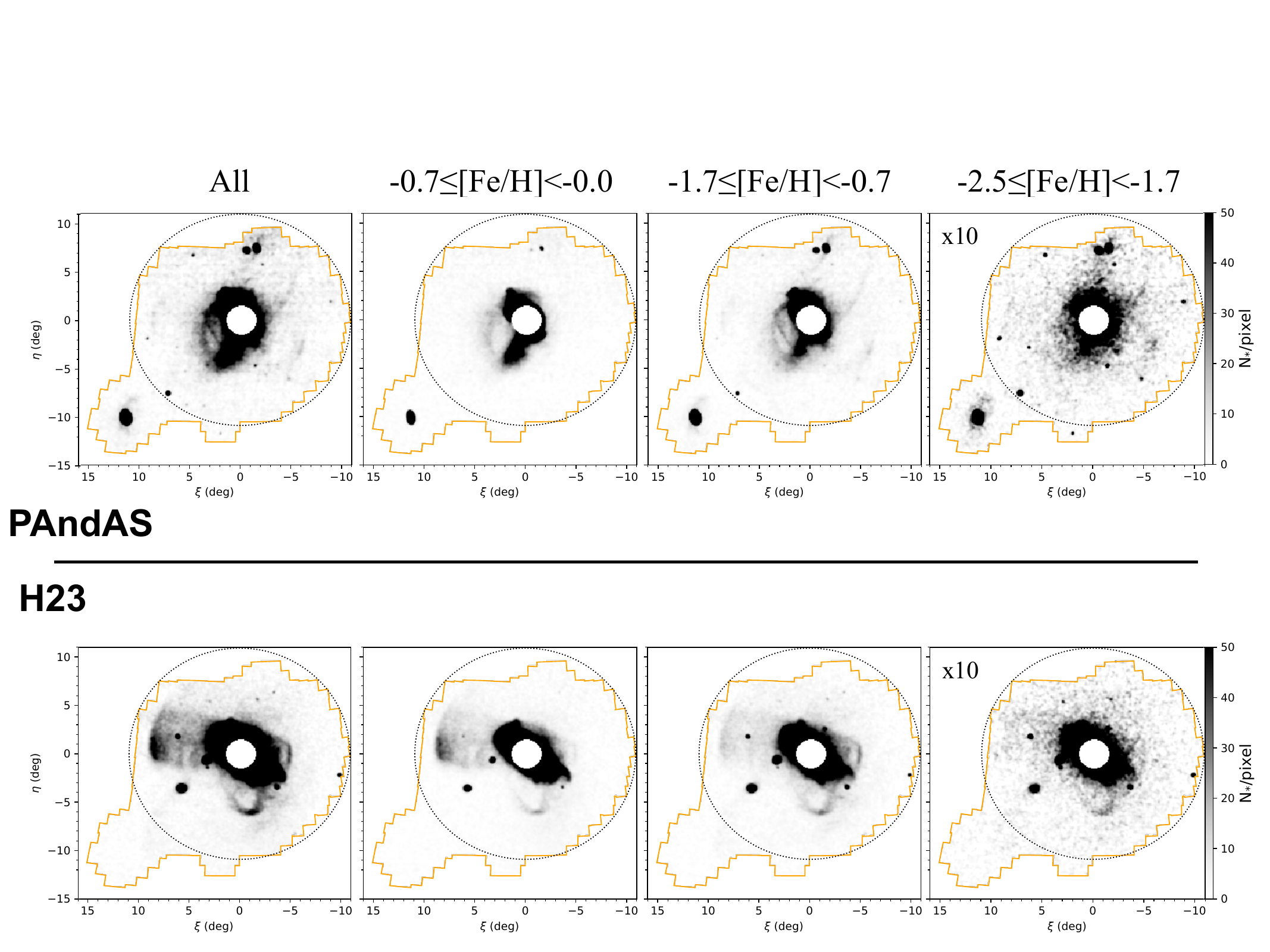}
   \caption{Map of stars in PAndAS (top panels) and in the mock of halo H23 (bottom panels) for different ranges of photometric metallicities. The density has been multiplied by 10 for the lower metallicities panel (right-hand panel) for better visibility. The contamination from the foreground Milky Way dwarfs and from the unresolved background galaxies has been removed statistically, assuming the contamination model of \citetalias{martin_2013}. The dotted circle represents a projected distance of 150 kpc.}
\label{map_den}
\end{figure*}

\begin{figure*}
\centering
  \includegraphics[angle=0, clip ,width=17.5cm]{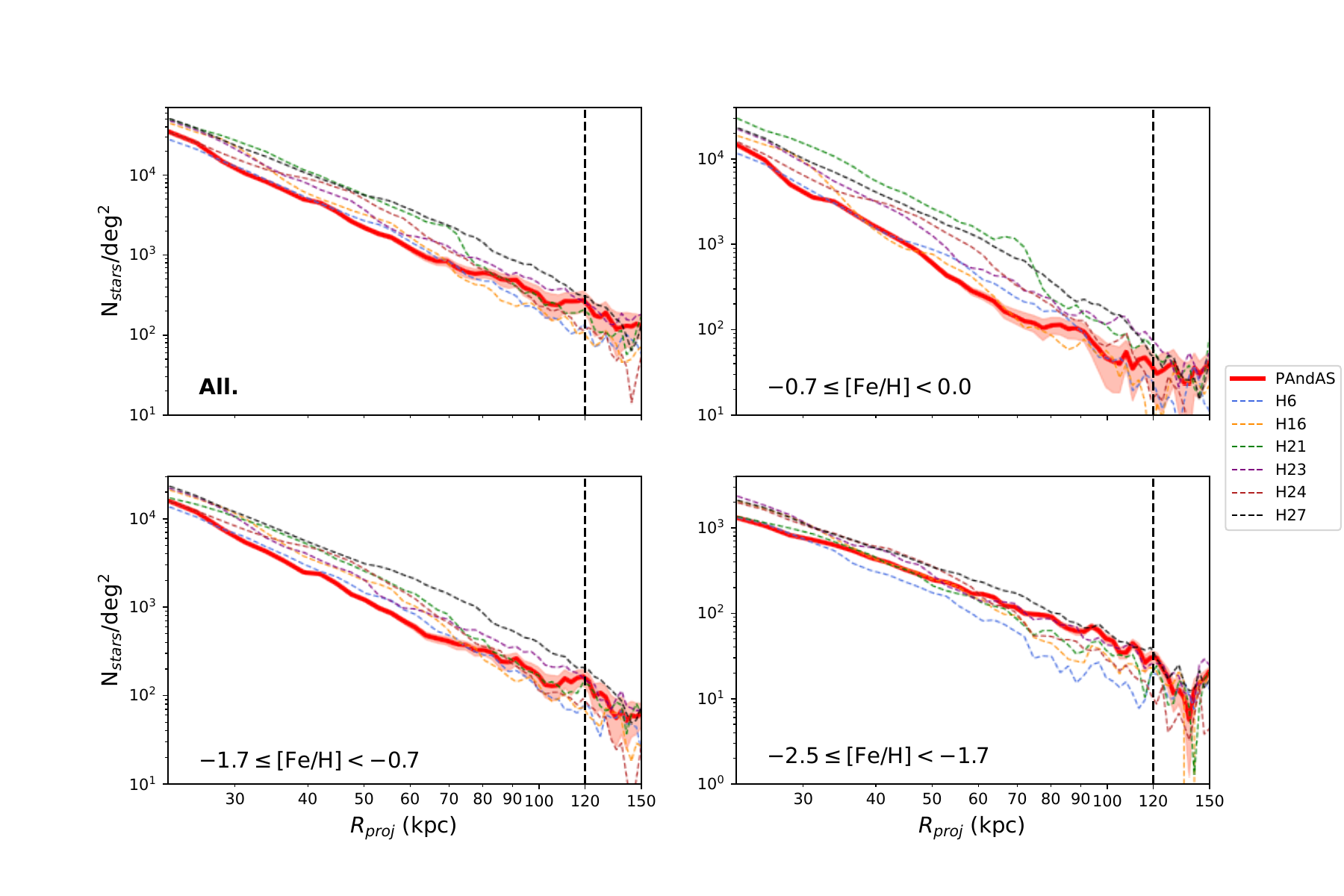}
   \caption{Profile of the median surface density of the different halos in four ranges of metallicities: all metallicities, metal-poor ($-2.5\leq$[Fe/H]<$-1.7$), intermediate metallicity ($-1.7\leq$[Fe/H]<$-0.7$), and metal-rich ($-0.7\leq$[Fe/H]<$0.0$). The scales are similar for all panels, except for the metal-poor range, where the scale is divided by ten for easy comparison with the other panels. In each panel, the red shaded area show the variation in the observed surface profile that we derived in PAndAS accounting for the 1-$\sigma$ uncertainties on the correction factor applied to the \citetalias{martin_2013} decontamination model, as found by \citetalias{ibata_2014b} ($\mathcal{F}=0.93\pm{0.03}$).}
\label{profile_all}
\end{figure*}

Despite this selection, a large fraction of PAndAS stars present in this region of the CMD are actually contaminant objects. The source of this contamination is mostly due to the foreground Milky Way dwarf stars and to unresolved background galaxies that appear as point sources. Therefore, in our attempt to produce realistic ``observed'' stellar halo mocks, it is important to add this contamination component, especially since the density of contaminants is not constant over the survey because of the increasingly dense MW disk towards the North. The addition of this component is extremely important, in order to treat the observations and the "realistic" mocks with the same methods, since most analyses will remove, or at least take into account, the contamination to enhance the signal produced by the stars from the stellar halo of M31. 

The model of the contamination used by the pipeline is based on the 4-dimensional (spatial and color-magnitude) model of \citetalias{martin_2013}, for which the density of contaminant objects at a position ($\xi, \eta$) and at a given color and magnitude ($g-i,i$) follow an exponential such that

\begin{equation}
\Sigma_{(g-i,i)}(\xi, \eta)= \exp(\alpha_{(g-i,i)}\ \xi + \beta_{(g-i,i)}\ \eta + \gamma_{(g-i,i)}).
\label{eq_conta}
\end{equation}

\noindent The $\alpha_{(g-i,i)}$, $\beta_{(g-i,i)}$, and $\gamma_{(g-i,i)}$ parameters were determined for any location $(g-i,i)$ in the CMD from a region outside $\sim 120$ kpc from Andromeda's center, assuming that the density of objects in this external region is produced uniquely by contamination (the reader is referred to \citetalias{martin_2013} for a detailed description of the model). The parameters $\alpha_{(g-i,i)}$, $\beta_{(g-i,i)}$, $\gamma_{(g-i,i)}$ used here are slightly different than the ones of \citetalias{martin_2013} since we now use the new public reduction of PAndAS presented in \citet{mcconnachie_2018}.

Practically, and following \citetalias{martin_2013}, the number of contaminant objects per spatial pixel of 15 $\times$ 15 arcmin$^2$, $N_j$, is computed using Eq.(\ref{eq_conta}) for the value of $\alpha_{(g-i,i)}$, $\beta_{(g-i,i)}$ and $\gamma_{(g-i,i)}$ and integrating the model counts in the CMD region delineated by the M31 RGB selection box mentioned above. This number is multiplied by $\mathcal{F}=0.93$, since \citetalias{ibata_2014b} show that $\simeq 7\%$ of the objects of the external region over which the contamination model was constructed, and which are in a color-magnitude region similar to the M31 RGB box, are actually stars from the halo of M31. Moreover, to incorporate statistical fluctuations to the contamination model, the actual number of contaminants for each pixel is drawn randomly, assuming a Poissonian distribution centered on $N_j$. The spatial locations of the $N_j$ contaminant particles of a given pixel are then randomly distributed following a uniform distribution over the pixel. 

The $g$ and $i$ magnitudes of the contaminant particles are then randomly distributed following the probability distribution function (PDF) of the contaminant CMD at the location of the spatial pixel of the ``stars'' (see \citetalias{martin_2013}). It is important to note here that the magnitudes given by the model are already corrected for the extinction. The photometric uncertainties of each contaminant ``star'' are then computed using Eq.(\ref{eq_uncert}). Finally, only the contaminant ``stars'' that are contained in the M31 RGB box are kept. This removes the few particles from pixels cut by the M31 RGB box that are outside this box. Depending on the science goal with the mocks, the M31 RGB box criterion can be removed from the overall pipeline (i.e. for particles from the stellar mock and for contaminants). However, in that case the number of contaminants will be slightly overestimated, as mentioned earlier, but by less than $7\%$, since this number was obtained in the M31 RGB box, which contains most of the M31 halo stars.

The final distribution of the ``stars'' for mock H23, including ``stars'' from the stellar mock and from the contamination model, is shown in panel (e) of Figure \ref{steps}.

The density of stars for a full realization of this halo is presented in Figure \ref{density_RGB}. The lower left panel shows the total density of ``stars'' that are in the color-magnitude space delimited by the M31 RGB box. This includes the ``stars'' from the stellar mock, whose distribution is shown in the upper left panel, and contaminants, shown in the upper right panel. While the details of the substructures are of course different because of the difference in the accretion history of the simulated galaxy and Andromeda, we note qualitative consistency between the mock and the PAndAS data.

\begin{table*}
 \centering
  \caption{Table of parameters of the stellar halos from {\sc Auriga}. The values are from \citet{monachesi_2019}.}
   \label{cat_sim}
  \begin{tabular}{@{}c|c|c|c|c|c@{}}
  \hline
   No. sim. & R$_{\rm vir}$ (kpc) & M$_{\rm vir}$ ($10^{12}$ M$_\odot$) & M$_{\rm stellar}$ ($10^{10}$  M$_\odot$) & M$_{\rm acc}$ ($10^{10}$  M$_\odot$) & M$_{\rm in-situ}$ ($10^{10}$  M$_\odot$)   \\
     \hline
   H6 & 213.82 & 1.04 & 5.41 & 0.38 & 0.64 \\ 
   H16 & 241.48 & 1.50 & 7.01 & 0.50 & 0.85 \\
   H21 & 238.64 & 1.45 & 8.65 & 1.17 & 1.03 \\
   H23 & 245.27 & 1.58 & 9.80 & 0.90 & 0.79 \\
   H24 & 240.85 & 1.49 & 7.66 & 0.64 & 0.74 \\
   H27 & 253.80 & 1.75 & 10.27 & 0.85 & 1.02 \\
    \hline
\end{tabular}
\end{table*}

\section{Results} \label{sec_results}

The \textsc{Auriga2PAndAS} pipeline is applied to the 6 simulated galaxies selected by \citet{grand_2018a}, H6, H16, H21, H23, H24, and H27, whose parameters are listed in Table~\ref{cat_sim}. These galaxies have virial masses\footnote{Here, we define the virial radius as the radius where the mean density is equal to 200 times the critical density of the universe.} ranging from $1.0$ to $1.8 \times 10^{12}$ M$_\odot$ and have virial radii of $\sim 240$ kpc. They cover the range of virial masses found for M31 using different tracers \citep[$0.8$ to $ 2.0 \times 10^{12}$ M$_\odot$][]{chemin_2009,watkins_2010,tollerud_2012,fardal_2013,veljanoski_2014,penarrubia_2016,kafle_2018}. The stellar mass of the simulated halos inside their virial radii varies from 6.6 to $10.3 \times 10^{10}$ M$_\odot$. Their stellar mass inside the inner 30 kpc range from 6.2 to 9.6 $\times 10^{10}$ M$_\odot$, which is slightly less massive than the $10.3^{+2.3}_{-1.7} \times 10^{10}$ M$_\odot$ found by \citet{sick_2015} for M31 inside this radius. However, in the outer halo (>30 kpc), the simulations have stellar mass of $0.8-1.0 \times 10^{10}$ M$_\odot$, similar to the  the mass of the stellar halo of M31 found by \citepalias{ibata_2014b} ($\simeq 1\times 10^{10}$ M$_\odot$. The only exception is the simulation H6, where the stellar mass outside 30 kpc is smaller than for the other simulated galaxies (0.3 $\times 10^{10}$ M$_\odot$), but is the consequence that this galaxy is the less massive of this sample.

Prior to any analysis and following \citetalias{martin_2013} and \citetalias{ibata_2014b}, we fill the holes of the PAndAS coverage present in the observations and included in the mocks by the pipeline. To do so, we duplicate stars from neighboring regions, both for the mocks and for the observations. To select the duplicated stars, the observations of the mocks are shifted by $0.15 \degr$ along both the $\xi$ and $\eta$ axis and the stars falling in the gaps are kept.

The fully processed mocks that are directly comparable to the PAndAS observations are distributed with this publication. A full description of the catalogues is provided in Appendix \ref{annexe} and the data themselves are accessible on the journal's website.

\subsection{Qualitative description of the simulations} \label{sec_desc}

Figure \ref{MF_all} shows a mapping of the structures present in the stellar halo of M31 and in the 6 mocks. Each of these maps is a red-green-blue image, made from the combination of matched filter (MF) maps whose filters are CMD models of old RGB stars at the distance of M31, convolved by the photometric uncertainties. The background model of the MF technique is the contamination model of \citetalias{martin_2013}. The blue image corresponds to a signal from RGB stars with a metallicity of $[\mbox{Fe/H}]=-2.3$, the green to $[\mbox{Fe/H}]=-1.4$, and the red to $[\mbox{Fe/H}]=-0.7$. The maps are made of spatial pixels of $3' \times 3'$ and have been smoothed by a Gaussian kernel of one pixel width. This is similar to the MF technique used by \citetalias{martin_2013} to produce the map of their Figure~2. The holes visible at the very center of the maps for the mocks are caused by the cut made to mask the stars within a three-dimensional radius of 20 kpc, as described in Section \ref{method}.

A visual inspection of these maps shows that the simulated stellar halos present numerous, well-defined substructures on a similar scale to those observed around M31 (central panel). In the central 50 kpc, the simulations and observations are very similar, with an inner stellar halo dominated by metal rich-stars ([Fe/H] $\sim -0.7$) and a small number of identifiable structures. Most of the mocks, with the exception of H6 and H24, have a large and metal-rich stream or shell, sign of a recent, or on-going, massive accretion. These structures are similar to the Giant stream\footnote{We follow \citet{lewis_2013} and \citet{mcconnachie_2018} for the nomenclature of the halo structures of M31.} \citep{ibata_2001a} visible in PAndAS that is the consequence of the accretion of a galaxy with a mass similar to the Large Magellanic Cloud $2-3$ Gyrs ago \citep{fardal_2013}. 

On average, the mock halos are redder, and so more metal-rich, than the observations. The halo of M31 presents more structures, such as stellar streams or clouds, at intermediate/low metallicity than the simulations. It is important to note here that the absence of a galaxy similar to M33 in some of the simulations is not surprising. Indeed, M33 would not be visible in the PAndAS footprint if it were located at the same projected distance but at a different angle around M31, as it is likely to be the case in the mocks. Furthermore, not all simulated halos presently have a satellite that is as massive as M33.

For each of the halos, we compute the surface density of RGB stars in spatial pixels of $0.1 \degr \times 0.1 \degr$ for different ranges of metallicities, in the same way as for the observations. In each of these pixels, the number of contaminant objects have been statistically removed assuming the model of \citetalias{martin_2013}. To compare directly the mocks to the observations, the metallicities have been derived photometrically for each star by comparing their colors and magnitudes to the Dartmouth isochrones \citep{dotter_2008} assuming that the stars are at the distance of M31, have an age of 10 Gyrs and an alpha enhancement of [$\alpha$/Fe]$=+0.2$, typical of the populations in the halo of M31/MW \citep[i.e.][]{helmi_2008,kilic_2019}. The surface density maps in four ranges of metallicities (all, metal-poor, intermediate metallicity and metal-rich stars) for the halo of M31 and for mock H$23$ are shown on Figure \ref{map_den}. The information this figure contains is similar to that visible in Figure \ref{MF_all}, but it is easier to see the contribution of each structure in the different ranges of metallicity. For instance, it clearly shows that H23 has a similar number of dwarf galaxies at intermediate metallicities compared to M31, but has only about half its number of metal-poor satellite galaxies similar to AndXXI or AndXXIII ($M_V \sim -9.5$). This characteristic is present in all mocks, especially for H6 and H21, for which the metal-poor galaxies are far less numerous but also more centrally distributed than observed around M31.

\begin{table*}
\centering
  \caption{Slopes of the power-law fit to the projected density profile for the different halo in the four different ranges of metallicity. For the mocks, the middle column lists the slopes of the overall populations in the halos (accreted and in-situ) and the right column lists the slopes of the accreted population only.}
  \label{table_slope}
  \begin{tabular}{@{}lr|c|c@{}}
  \hline
   & {\bf [Fe/H] range} & {\bf All}  & {\bf Accreted} \\
  \hline
{\bf PAndAS~~} & All                & $-3.39^{\pm{0.01}}$ & -\\
 & $-2.5\leq$[Fe/H]<$-1.7$ & $-2.23^{\pm{0.06}}$ & -\\
 & $-1.7\leq$[Fe/H]<$-0.7$ & $-3.24^{\pm{0.01}}$ & -\\
 & $-0.7\leq$[Fe/H]<$\ 0.0$ & $-4.04^{\pm{0.02}}$ & -\\
\hline
\hline
{\bf H6} & All              & $-3.22^{\pm{0.01}}$ & $-3.13^{\pm{0.01}}$  \\
 & $-2.5\leq$[Fe/H]<$-1.7$  & $-2.95^{\pm{0.09}}$ & $-2.88^{\pm{0.10}}$\\
 & $-1.7\leq$[Fe/H]<$-0.7$  & $-3.07^{\pm{0.01}}$ & $-3.06^{\pm{0.01}}$\\
 & $-0.7\leq$[Fe/H]<$\ 0.0$ & $-3.57^{\pm{0.02}}$ & $-3.39^{\pm{0.02}}$\\
\hline
{\bf H16} & All             & $-3.67^{\pm{0.01}}$ & $-3.49^{\pm{0.01}}$  \\
 & $-2.5\leq$[Fe/H]<$-1.7$  & $-2.95^{\pm{0.06}}$ & $-2.84^{\pm{0.07}}$\\
 & $-1.7\leq$[Fe/H]<$-0.7$  & $-3.38^{\pm{0.01}}$ & $-3.30^{\pm{0.01}}$\\
 & $-0.7\leq$[Fe/H]<$\ 0.0$ & $-4.39^{\pm{0.02}}$ & $-4.09^{\pm{0.02}}$\\
\hline
{\bf H21} & All             & $-3.20^{\pm{0.01}}$ & $-2.98^{\pm{0.01}}$  \\
 & $-2.5\leq$[Fe/H]<$-1.7$  & $-2.57^{\pm{0.06}}$ & $-2.51^{\pm{0.07}}$\\
 & $-1.7\leq$[Fe/H]<$-0.7$  & $-2.95^{\pm{0.01}}$ & $-2.88^{\pm{0.01}}$\\
 & $-0.7\leq$[Fe/H]<$\ 0.0$ & $-3.48^{\pm{0.01}}$ & $-3.19^{\pm{0.01}}$\\
\hline
{\bf H23} & All             & $-3.35^{\pm{0.01}}$ & $-3.26^{\pm{0.01}}$  \\
 & $-2.5\leq$[Fe/H]<$-1.7$  & $-2.76^{\pm{0.05}}$ & $-2.73^{\pm{0.07}}$\\
 & $-1.7\leq$[Fe/H]<$-0.7$  & $-3.17^{\pm{0.01}}$ & $-3.21^{\pm{0.01}}$\\
 & $-0.7\leq$[Fe/H]<$\ 0.0$ & $-3.76^{\pm{0.01}}$ & $-3.50^{\pm{0.01}}$\\
\hline
{\bf H24} & All             & $-3.05^{\pm{0.01}}$ & $-2.80^{\pm{0.01}}$  \\
 & $-2.5\leq$[Fe/H]<$-1.7$  & $-2.78^{\pm{0.05}}$ & $-2.71^{\pm{0.06}}$\\
 & $-1.7\leq$[Fe/H]<$-0.7$  & $-2.91^{\pm{0.01}}$ & $-2.74^{\pm{0.01}}$\\
 & $-0.7\leq$[Fe/H]<$\ 0.0$ & $-3.38^{\pm{0.01}}$ & $-3.03^{\pm{0.01}}$\\
\hline
{\bf H27} & All             & $-2.91^{\pm{0.01}}$ & $-2.70^{\pm{0.01}}$  \\
 & $-2.5\leq$[Fe/H]<$-1.7$  & $-2.40^{\pm{0.04}}$ & $-2.37^{\pm{0.05}}$\\
 & $-1.7\leq$[Fe/H]<$-0.7$  & $-2.69^{\pm{0.01}}$ & $-2.61^{\pm{0.01}}$\\
 & $-0.7\leq$[Fe/H]<$\ 0.0$ & $-3.28^{\pm{0.01}}$ & $-2.96^{\pm{0.02}}$\\
\hline
\end{tabular}
\end{table*}

\begin{figure*}
\centering
  \includegraphics[angle=0, clip ,width=17.5cm]{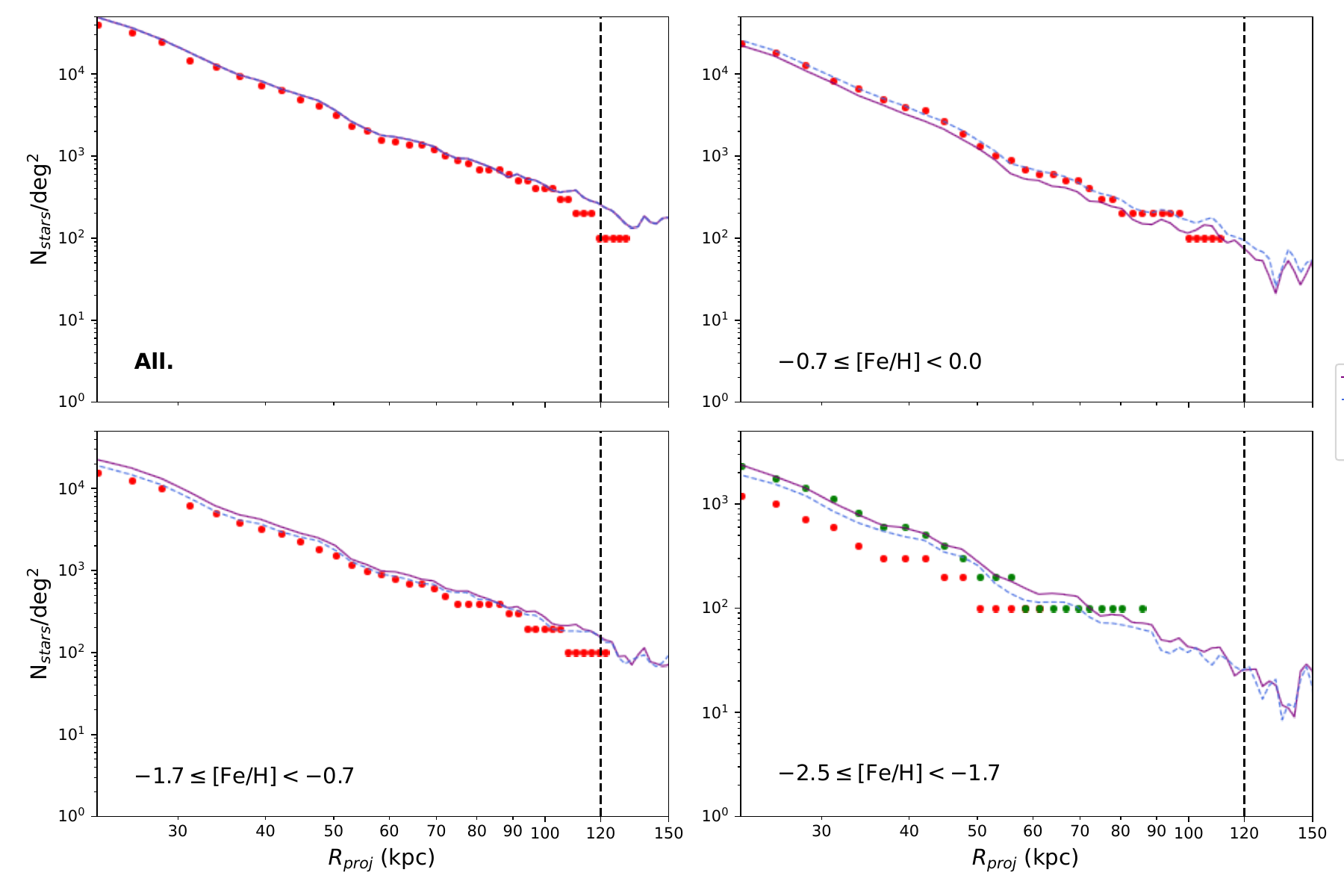}
   \caption{Comparaison of the median surface density profile of mock H23 with the photometric metallicities determined from the Dartmouth isochrones used for figure \ref{profile_all} (purple), the 8-Gyr old Padova isochrones (blue) and with the original ``true'' metallicities of the simulations' stellar particles (red dots). The profile shown by the green dots corresponds to the profile of the stellar particles in the range $-2.5\leq$[Fe/H]$_\mathrm{sim}<-1.5$. The plateau at 100 stars/deg$^2$ for this profile is due to the fact that we take the median of the number of actual stellar particles per pixels of $0.1$ deg $\times 0.1$ deg, which are by definition an integer. The profiles determined by the photometric metallicities do not have this behavior because the number of foreground stars per pixel is removed statistically.}
\label{profile_truth}
\end{figure*}

We use these surface-density maps to derive the median surface density radial profile of RGB stars in the four ranges of metallicities for all the halos. The median profile is favored over the mean one since it limits the contribution from compact substructures, such as satellite galaxies. The resulting profiles are shown in Figure \ref{profile_all} and fitted with a single power-law profile in the range of projected distances of $23 - 120$ kpc ($9 \degr$ at the distance of M31). The resulting power-law slope values are listed on Table \ref{table_slope}. We purposefully avoid the region beyond 120 kpc since it was used to construct the contamination model of \citetalias{martin_2013}, leading to profiles that could be systematically biased at these radii, despite the correction mentioned in Section \ref{sel_rgb}.

When considering the surface density profiles for the full metallicity range, all mocks apart from H27 have profile slopes relatively similar to M31. However, most of the mocks are more centrally populated than M31, with higher surface densities, up to a factor $\sim 4$, until $\sim 75$ kpc. Beyond $\sim 100$ kpc, most of the mocks are less densely populated than the observed M31 halo. It is also interesting to note here that the profile of H21 shows a clear drop at $\sim 75$ kpc, also visible in the metal-rich range. This is caused by the 2 major accretions that are ongoing and largely dominate the rest of the halo population up to $\sim 75$ kpc. 

By comparing the density profiles in the different metallicity intervals, it is clear that the global surface density profiles of the stellar halos are dominated by metal-rich and intermediate metallicity stars, both in M31 and in the mocks. However, the difference of density between the observations and the simulations are most prominent for metal-rich stars ($-0.7\leq$[Fe/H]$<0.0$), for which the density of stars in the mocks is higher than the observations at all the radii. In this metallicity range, H16 is the exception, with surface density profiles that are similar to the ones observed for M31 beyond 30 kpc. At intermediate metallicities ($-1.7\leq$[Fe/H]$<-0.7$), the differences are smaller than in the metal-rich regime, with profile of the mocks being, on average, $\simeq$ 1.5 more populated that of M31 at intermediate metallicities, against $\simeq$ 2.2 at metal-rich metallicities. Overall, the mocks are still more densely populated than observed until a projected radius of $\sim 100$ kpc. Beyond that distance, the density in most of the simulated halos drops below the observed density.

For the metal-poor star selection ($-2.5\leq$[Fe/H]$<-1.7$), half of the mocks are $\sim 1.7$ times more populated than M31. The other half have a similar inner density than observed in M31. However, in this metallicity range, the surface density profile of the mocks is systematically steeper than observed, and for most of them, the distant halo ($>60$ kpc) shows a clear deficit of metal-poor stars compared to the observation. H27 is an exception compared to the other simulations since, for all metallicity intervals, the surface density of the mock is always higher than observed in M31 at any radius. 

Given these observations, we can conclude that the halos of the simulated galaxies are, in general, more populated than the M31 halo, especially in the central region. In particular, the simulations are more populated by metal-rich stars than observed around M31 at any radius. Moreover, the metal-poor stars in the mocks are more centrally concentrated and are less extended than observed by PAndAS around M31.

\begin{figure*}
\centering
  \includegraphics[angle=0, clip ,width=17.5cm]{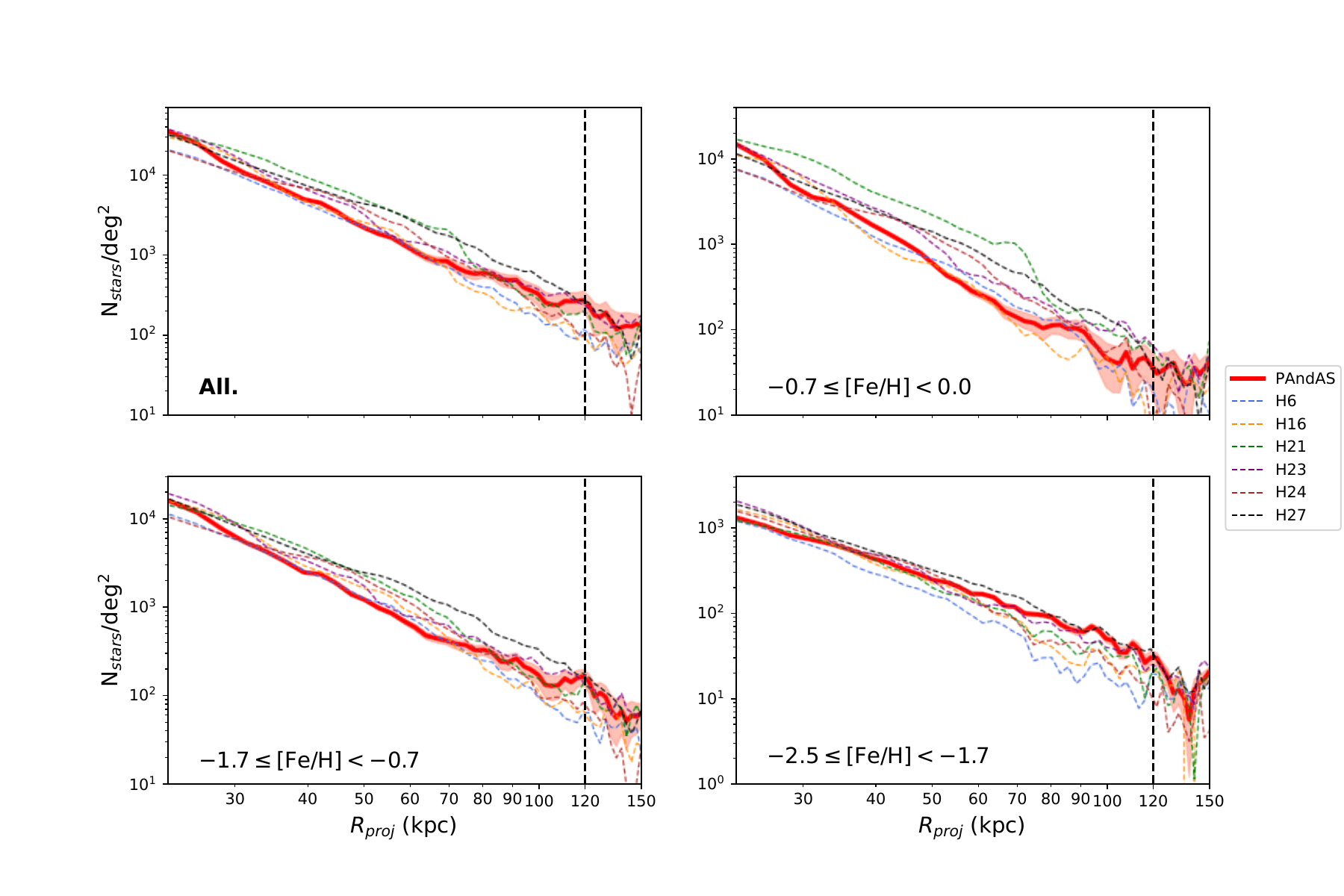}
   \caption{Median surface profile, similar to Figure \ref{profile_all} but just for the accreted ``stars.''}
\label{profile_accreted}
\end{figure*}

\subsection{Precision of the photometric metallicities} \label{sec_photomet}

In the previous section, the analysis is partly based on photometric metallicities derived using the Dartmouth isochrones, while the Padova isochrones are used to give an absolute magnitude to the ``stars'' in the simulations. In this section, we show that the choice of the set of isochrones has a negligible impact on our analysis, and that this choice does not alter the conclusions that we draw from the analysis of the mocks.

For the comparison between the mocks and the observations, the Dartmouth isochrones were preferred over the Padova ones for mainly two reasons. The first is that, historically, the previous works done by the PAndAS collaboration were used mainly the Dartmouth isochrones \citep[e.g.][]{mcconnachie_2009,collins_2010,martin_2013,ibata_2014b,mcconnachie_2018}. Thus, using the same set of isochrones facilitates a direct comparison with these previous analyses. The second is that using a set of isochrones that is independent from the set used to generate the stellar particles in the simulations removes any suspicion of possible biases in the analysis.

Figure \ref{profile_truth} displays the median surface density profile for mock H23 in the four previous ranges of metallicities, for the case where the photometric metallicities are derived using the 10 Gyr old Dartmouth isochones with [$\alpha$/Fe]=+0.2 (purple lines), like in the previous section, and with the Padova isochrones (blue dashed lines) corresponding to a population of 8 Gyr, the median age of the ``stars'' in this specific halo. This figure also include the median surface profile of the un-contaminated ``stars'' (i.e. that does not include the ``stars'' from the foreground model), using the metallicities directly provided by the simulations (red dots). These different profiles are very similar in all metallicities ranges, except for the metal-poor regime. In this range, the two profiles obtain using the photometric metallicities are still very similar to each other but are systematically higher than the``true'' profile by a factor $\sim2$. This is explained by the fact that isochrones of metal-poor stars are very close to each other in a CMD. Thus, with the photometric uncertainties and the intrinsic scatter of a stellar population, metal-poor stars overlap each other in a CMD, leading to a lower accuracy of the photometric metallicities in that regime. Actually, it seems that a photometric metallicity range of $-2.5\leq$[Fe/H]$_\mathrm{photo}<-1.7$ corresponds to a range of $-2.5\leq$[Fe/H]$_\mathrm{sim}<-1.5$ for the metallicities of the mocks, for which the profiles are represented by the green dots on the lower-right panel. 

However, it is important to keep in mind that our analysis is based on relative metallicities, with four broad ranges. Therefore, whether a given ``star'' has [Fe/H]=-1.7 or [Fe/H]=-1.5 does not significantly impact the fact that metal-poor stars are less numerous than the intermediate or metal-rich ones. In addition, we perform the exact same analysis on the PAndAS observations and on the {\sc Auriga} mocks. The conclusions that we draw are therefore based on a relative comparison of these results and do not rely of the absolute metallicity values.

\subsection{Comparison of the in-situ/accreted components} \label{sec_insitu}

It has been shown with semi-analytical and hydrodynamical simulations of L$\star$-galaxies that their stellar halo can be decomposed into two populations with very different origins, the {\it in-situ} and the accreted components \citep[also referred to as the {\it ex-situ} component in the literature, e.g.][]{bullock_2005,cooper_2010,font_2011,purcell_2011,pillepich_2014,monachesi_2019}. The accreted component is made of stars initially hosted by dwarf galaxies and globular clusters that have since been disrupted by tidal effects and and which have deposited their stars into the main halo. The {\it in-situ} component is made of stars from the galactic disk that have been ejected into the halo due to the dynamical heating produced by the interactions with sub-halos, galaxies or massive molecular clouds, and by stars that were born in the halo of the proto-galaxy.

To differentiate the contribution of these two populations in the mocks, we redo the analysis done in Section \ref{sec_desc}, but select only the {\it accreted} ``stars'' in the mocks. Mock ``stars'' are classified as {\it accreted} or {\it in-situ} according to their parent stellar particle, following the definition presented in Section 2.3 of \citet{monachesi_2019}. In this definition, stellar particles bound to the host galaxy at birth are considered as {\it in-situ} stars.

As visible in Figure \ref{profile_accreted}, considering only the accreted stars, the surface density profiles of the simulated galaxies are generally in better agreement with the profile of M31. In all the different metallicity intervals, the shape of the surface density profiles of the accreted ``stars'' alone are closer to the profiles observed in M31 than if we consider the full stellar halo populations (in situ and ex situ), especially for simulation H23. The strongest deviations between the surface density profiles of the accreted populations and the overall population (accreted+{\it in-situ}) are visible with the most metal-rich stars, followed by the intermediate metallicities range. This is not surprising since the accreted stars are on average more metal-poor than the stars formed {\it in-situ}, as seen in the Milky Way \citep[i.e.][]{carollo_2007,carollo_2010,belokurov_2019} and in different suites of cosmological simulations \citep[i.e.][]{purcell_2011,pillepich_2014,monachesi_2019}.

\begin{figure*}
\centering
  \includegraphics[angle=0, clip ,width=17.5cm]{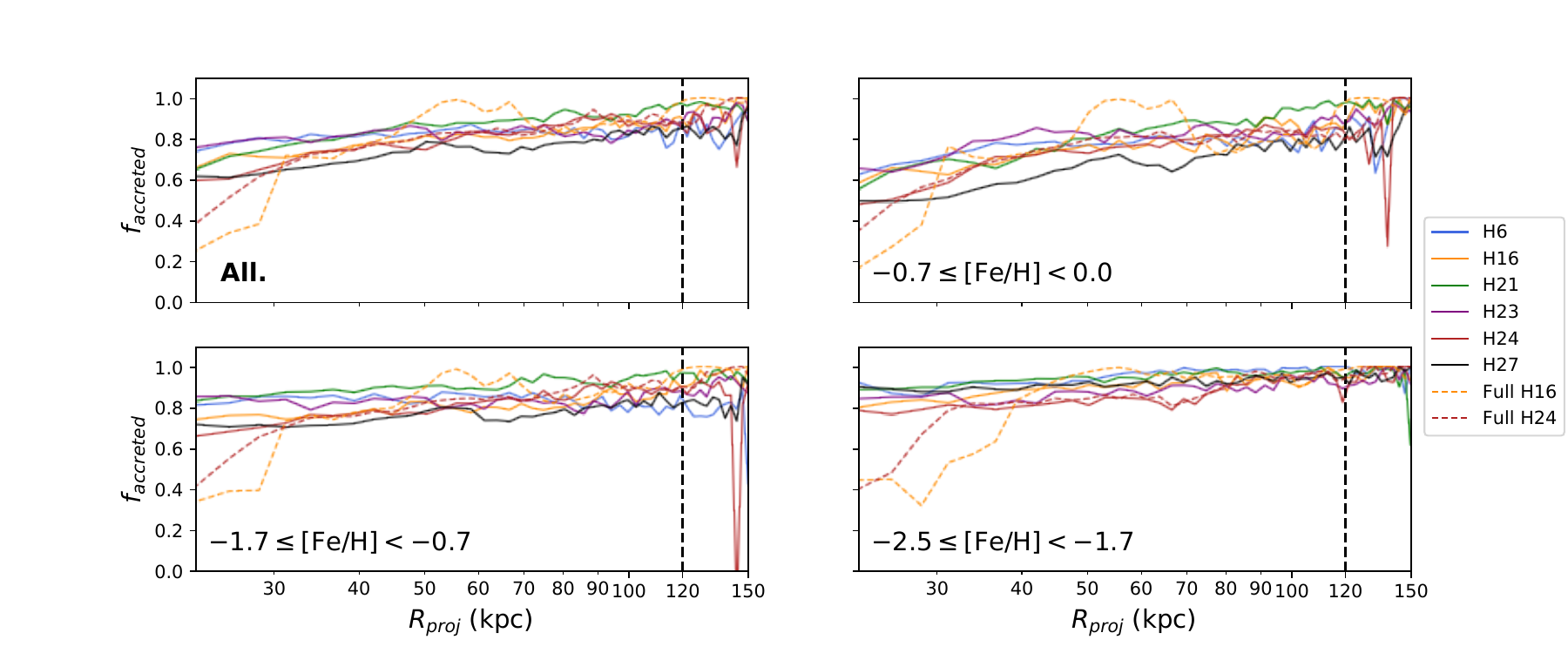}
   \caption{Projected profile of the fraction of accreted ``stars'' in the mocks.}
\label{ratio}
\end{figure*}

As visible on Figure \ref{ratio}, the accreted stars make up the majority of the stellar halos (at least beyond 23 kpc) for all metallicities ranges, as noted by \citet{monachesi_2019}. However, the radial profile of the fraction of accreted stars is very different in the different metallicity intervals. Indeed, for metal-rich stars, $\simeq 50-65 \%$ of them are accreted at 23 kpc and this fraction increases slowly to reach $\simeq 80-90\%$ at 120 kpc, while for the intermediate and metal-poor stars the fraction of accreted stars is almost constant across all radii, accounting for $\sim 70-80 \%$ and $\sim 90 \%$ of the stars in these respective ranges. This confirms that the {\it in-situ} component is mostly composed of centrally concentrated metal-rich stars, similar to what is observed in the Milky Way or in different simulations \citep[i.e.][]{purcell_2010,cooper_2015,pillepich_2015,monachesi_2019,sanderson_2018,belokurov_2019}. Note that the ratio shown in this figure corresponds to the ratio of stars from the smooth halo component (i.e. not in sub-structures). Taking into account the overall population, the ratios are similar for all mocks, except for H16 and H24 (referred as the Full populaiton of Figure \ref{ratio}), for which the fraction of in-situ stars is significantly higher, up to $\sim 35$ and $25$, kpc respectively. This is caused by the presence of very extended galactic disk in these simulations, as noticed by \citet{grand_2018}. However, the goal of this section is to compare the fraction of accreted stars in the halos of the different simulated galaxies and so we purposefully do not take into account the stars in the disk or those present in sub-structures.

By comparing the surface density profile observed in M31 in the different ranges of metallicity and the profiles of the accreted and of the accreted+{\it in-situ} populations in the mocks, it seems that the mocks show a density of {\it in-situ} stars that is about twice as large as observed in M31. Indeed, by reducing roughly by a factor 2 the contribution of the {\it in-situ} component, the surface density profiles will be closer to the profile observed in PAndAS, especially for H16, H21 and H23. This is most visible when comparing the accreted and overall metal-rich surface density profile in the inner-halo ($<30$ kpc). This conclusion is in agreement with the observation of \citet{monachesi_2019} using the {\sc Auriga} simulations, and similar results have recently been found by \citet{merritt_2020} comparing the {\sc Illustris TNG100} simulations \citep{springel_2018,pillepich_2018} to the Dragonfly Nearby Galaxies Survey \citep{merritt_2016}.

However, even by reducing the number of the stars formed {\it in-situ} by a factor of 2, most of the mocks have halos that are overall more populated than M31's and their profiles present a great diversity. Therefore, outside the inner halo, where the majority of the {\it in-situ} stars are located, this scenario is not the only explanation for the observed difference between M31 and the simulated galaxies. Because galaxies like H$23$ have a very similar profile to the observed one, this overpopulation of most of the galaxies and the diversity of profiles is possibly driven by the stochastic process on the merger histories of each galaxies.

From all the simulated halos analysed here, the halo H$23$ is the most representative of the halo of M31. Not only does this halo have a higher number of dwarf galaxies similar to AndXXI or AndXXIII than the other simulations, but its surface density profile is the closest to the profile constructed from PAndAS in all metallicities ranges, especially for the accreted ``stars''. However, as for all the simulated halos, the {\it in-situ} component is $\sim 2$ times more populated compared to M31.

\subsection{Analysis of the simulations} \label{sec_analyse}

As we have seen in the previous sections, the {\sc Auriga-PAndAS} mocks are, overall, reasonable approximations of the stellar halo of M31 and its diversity of structures. Qualitatively, the inner halo of the simulations are very similar to the inner halo of M31. Moreover, most of the halos are populated by a number of intermediate metallicity dwarf galaxies that are similar to what we see around M31. They cover a broad range of luminosities, from galaxies with a size and a metallicity similar to NGC 147 and NGC 185 (corresponding to a stellar mass of $\simeq 5\times 10^{7}$ M$_\odot$), to galaxies similar to AndXXI and AndXXIII (corresponding to stellar masses of $\sim 1\times 10^{6}$ M$_\odot$). However, in certain aspects, the mocks differ from what is observed around M31. Although the goal of the paper is not to analysis the reasons of the (overall) small differences between the simulations and the observations, we mention hereafter a few points that could be explored in future works. 

For instance, we can notice that the simulated stellar halos are on average more metal-rich than observed in M31. As explained in the previous section, this is likely a consequence of an over representation of the {\it in-situ} stars in the simulations, these stars being more metal-rich than the accreted ones. As mentioned by \citet{monachesi_2019}, this over-population of stars formed {\it in-situ} could be a consequence of the disks of the host galaxy, which, in the \textsc{Auriga} simulations, are typically larger than observed for M31 (see \cite{grand_2017} and \cite{monachesi_2019} for a discussion on the size of the disk). However, even with considering only the accreted stars, the simulations tends to be more metal rich than observed for all radii, which might be a consequence of the properties of the accreted dwarf galaxies that formed these structures but also, potentially, an under-quenching of the star formation feedback in the last billion years, especially in the most massive satellites, as proposed by \citet{monachesi_2019}.  Moreover, the surface density profile is lower that the majority of the simulation out to $\sim$ 70-80 kpc, even when considering only the accreted population. This is surprising given that the simulated galaxies have a similar or lower stellar mass than observed in M31. This could also be a consequence of the under-quenching of the star formation feedback, of the way in which the stellar particles are split into ``stars'' in the simulations, and/or of the IMF choice or the stellar populations models.

One of the other noticeable differences between the simulations and the observations of M31 is the relatively small number of metal-poor dwarf galaxies present around the simulated galaxies, compared to the numbers observed in PAndAS. Galaxies similar to And XIV or And XII are not visible at all in the simulations. \citet{simpson_2018} show that the number of dwarf galaxy {\sc Auriga} simulations converge down to  $2-3 \times 10^4$ M$_\odot$ for different resolutions. However, it is possible that the small number of metal-poor dwarf galaxies in the simulation  might be due to an incomplete sampling of the total phase-space of these galaxies, despite the fact that, with the \textsc{ICC} method, ``stars'' conserve the phase space of the initial stellar particle. It will be interesting to quantify these apparent differences by searching and characterizing the dwarf galaxies in the mocks in the same way they were found and characterized in PAndAS.

\section{Conclusions} \label{conclusion}
We presented a pipeline to transform the output of {\it top-down} {\sc Auriga} simulations as if they were observed by the CFHT telescope with the MegaCam instrument as part of the Pan-Andromeda Archaeological Survey, a survey that observed the stellar halos of M31 and M33 out to a projected radius of 150 kpc and 50 kpc respectively. Such a method is a basic requirement if we are to directly compare state-of-the-art cosmological simulations to exquisite observations of the Local Group in order to better constrain and refine the galaxy formation processes used in these simulations. The transformation of the simulations into ``observable'' mocks allows the use of the exact same tools, with all the assumptions that they include, on the observations and on the simulations.

We have performed a qualitative comparison between the simulated stellar halo of 6 L$\star$-galaxy mocks from the {\sc Auriga} simulations and the observed stellar halo of the Andromeda galaxy. We find that overall, the mock halos are similar to the halo of M31, with the presence of numerous structures similar to those observed. This is especially true in the inner halo ($<30$ kpc), where mocks and observations have very comparable structure and metallicity.  Moreover, most of the mocks present the sign of a recent massive accretion, similar to the Giant Stream observed in M31. Some of the mocks have a very similar density profile of accreted stars compared to observations, the variations between the different simulated galaxies being a consequence of the stochasticity of the hierarchical galaxy formation process. However, though it is challenging to conclude definitively with only 6 simulations to compare to, the mocks present some systematic differences with the observations. We see that the {\it in-situ} populations are overly represented by a factor $\simeq 2$, that the faintest dwarf galaxies visible in PAndAS are absent from the mocks, and that metal-poor structures like the NGC 147 stream are also absent. We also find that the metal-poor component of the different mocks is more concentrated than that observed in M31. We interpret these differences as a consequence of under-quenching by stellar feedback, increasing the average metallicity of the simulated galaxies, but also to a limitation of the resolution of the stellar particles in the simulations. It warns us against currently pushing the simulations into the very faint regimes probed by the PAndAS observations.

In future work, we will quantify the level of structure present in the simulations using the method presented in \citet{mcconnachie_2018} and we will compare it to the level of the different structures seen in PAndAS. Moreover, the distribution and the characteristics of the simulated satellite galaxies will be studied more in detail to compare to the distribution of the satellites of M31 and determine how faint the comparison is possible.

The {\sc Auriga2PAndAS} pipeline presented here (https://github.com/GFThomas/Auriga2PAndAS) and one realisation of each mock are publicly available online (https://wwwmpa.mpa-garching.mpg.de/auriga/gaiamock.html). The different columns of the catalogues are described in Appendix \ref{annexe}. We encourage the application of this pipeline to other state-of-the-art cosmological simulations such as EAGLE \citep{theeagleteam_2017}, Illustris TNG \citep{pillepich_2018a}, Latte/FIRE \citep{wetzel_2016,hopkins_2018}, or ARTEMIS \citep{font_2020}, so as to compare the results obtained from these simulations with the observations and between each other. Only then will be able to efficiently constrain the different parameters of the simulations, especially those relating to the complicated baryonic physics.

\acknowledgements
GT acknowledge support from the Agencia Estatal de Investigaci\'on (AEI) of the Ministerio de Ciencia e Innovaci\'on (MCINN) under grant with reference (FJC2018-037323-I).

FAG acknowledges financial support from CONICYT through the project FONDECYT Regular Nr. 1181264, and funding from the Max Planck Society through a Partner Group grant.
Based on observations obtained with MegaPrime/MegaCam, a joint project of CFHT and CEA/DAPNIA, at the Canada-France-Hawaii Telescope (CFHT) which is operated by the National Research Council (NRC) of Canada, the Institut National des Science de l'Univers of the Centre National de la Recherche Scientifique (CNRS) of France, and the University of Hawaii.

\bibliography{./biblio}

\appendix 
\section{Description of the online catalogue} \label{annexe}

\begin{table*}[h]
 \centering
  \caption{Description of each column in the online catalog available here: https://wwwmpa.mpa-garching.mpg.de/auriga/gaiamock.html.}
   \label{cat_online}
  \begin{tabular}{@{}l|l|l@{}}
  \hline
   No & Column name & Description \\
    \hline
0 & ID & Identifiant of the star from the Auriga simulations. -99 if from the background model \\
1 & RA & Right Ascension (deg) \\
2 & Dec & Declination (deg)\\
3 & xki & $\xi$ tangential coordinates centered on M31 (deg)\\
4 & eta & $\eta$ tangential coordinates centered on M31 (deg)\\
5 & rhelio & Heliocentric distance of each star (kpc). 0 if from the background model \\
6 & pmra & Proper motion in the right ascension direction  (mas/yr). 0 if from the background model \\
7 & pmdec & Proper motion in the declination direction  (mas/yr). 0 if from the background model \\
8 & Vrad & Heliocentric radial velocity (km/s). 0 if from the background model \\
9 & g & Apparent magnitude in the $g$-band \\
10 & dg & Uncertainty on the $g$-band magnitude \\
11 & g0 & Deredded $g$-band magnitude \\
12 & i & Apparent magnitude in the $i$-band \\
13 & di & Uncertainty on the $i$-band magnitude \\
14 & i0 & Deredded $i$-band magnitude \\
15 & EBV & Extinction from \citep{schlegel_1998} \\
16 & nb\_field & PAndAS field in which is the star (from 1 to 406) \\
17 & x & Galactocentric cartesian coordinate of the star from Auriga (kpc). 0 if from the background model\\
18 & y &  Galactocentric cartesian coordinate of the star from Auriga (kpc). 0 if from the background model\\
19 & z &  Galactocentric cartesian coordinate of the star from Auriga (kpc). 0 if from the background model \\
20 & Mg & Absolute magnitude in the $g$-band from the Auriga simulation. 0 if from the background model \\
21 & Mi &  Absolute magnitude in the $i$-band from the Auriga simulation. 0 if from the background model \\
22 & feh\_sim & Metallicity of the stars from the Auriga simulation.  0 if from the background model \\
23 & Acc & Flag the origin of the simulated stars (-1=formed in situ, 0=accreted, 1= formed in a satellite after infall).\\
 & & 0 for the stars from the background model. \\
\hline
\end{tabular}
\end{table*}

\end{document}